\pdfoutput=1
 \documentclass{article}
 \usepackage{graphicx}
\usepackage{amssymb}
\usepackage{amsmath}
\usepackage{amsmath}

\begin{document}
\unitlength 1cm
 
%\begin{frontmatter}



\begin{center}
{\Large {\bf Influence of the spatial resolution on fine-scale
features in DNS of turbulence generated by a single
square grid}}

\vspace{1cm}
{\large S. Laizet, J. Nedi\'c and J.C. Vassilicos}

\vspace{0.5cm}
{\it Turbulence, Mixing and Flow Control Group,\\ Department
  of Aeronautics, Imperial College London, \\ London SW7 2AZ, United
  Kingdom}

\vspace{0.5cm}
\end{center}


%\author{S. Laizet\corauthref{cor}}
%\ead{s.laizet@imperial.ac.uk}
%\author{J. Nedi\'c \thanksref{present}}
%\ead{j.nedic@imperial.ac.uk}
%\author{J.C. Vassilicos}
%\corauth[cor]{Corresponding author.}
%\ead{j.c.vassilicos@imperial.ac.uk}
%\thanks[present]{Present Address: Department of Mechanical Engineering,
%  University of Ottawa, Ottawa, Ontario, Canada K1N 6N5}

%\address{Turbulence, Mixing and Flow Control Group,\\ Department
%  of Aeronautics, Imperial College London, \\ London SW7 2AZ, United
%  Kingdom} 

\section*{Abstract}
We focus in this paper on the effect of the resolution of Direct
Numerical Simulations (DNS) on the spatio-temporal development of the
turbulence downstream of a single square grid. The aims of this study
are to validate our numerical approach by comparing experimental and
numerical one-point statistics downstream of a single square grid and
then investigate how the resolution is impacting the dynamics of the
flow. In particular, using the $Q$-$R$ diagram, we focus on the
interaction between the strain-rate and rotation tensors, the
symmetric and skew-symmetric parts of the velocity gradient tensor
respectively. We first show good agreement between our simulations and
hot-wire experiment for one-point statistics on the centreline of the
single square grid. Then, by analysing the shape of the $Q$-$R$
diagram for various streamwise locations, we evaluate the ability of
under-resolved DNS to capture the main features of the turbulence
downstream of the single square grid.
%\end{abstract}

%\begin{keyword}
%Turbulence \sep Direct Numerical Simulation 
%\end{keyword}
%
%\end{frontmatter}

\section{Introduction}

The most accurate approach to simulate a turbulent flow is to solve
the Navier-Stokes equations without averaging, extra modelling
assumptions and parameterisations (e.g. sub-grid) or approximations
other than numerical discretisations. This approach is called Direct
Numerical Simulation (DNS) and is the simplest approach conceptually
because all the motions of the flow are supposed to be resolved.
Because of the enormous range of scales in time and in space which
need to be resolved, DNS of turbulent flows can become very expensive
in terms of computational resources and is therefore often only used
for the understanding of the fundamental features of turbulence in
relatively simple flow configurations such as periodic homogeneous
isotropic turbulence or turbulent channel flows \cite{moin&mahesh98}.

With recent impressive developments in computer technology, it is now
possible to undertake DNS with a very large number of mesh nodes to
study turbulent flows at relatively high Reynolds numbers in more
complex configurations than homogeneous isotropic turbulence. One of
the main difficulties however is to determine the spatial resolution
of a DNS, as this choice is related to the range of scales that need
to be accurately represented. The resolution requirements are
obviously influenced by the numerical method used and the usefulness
of highly accurate numerical schemes for DNS is fully recognized, the
most spectacular gain being obtained with spectral methods based on
Fourier or Chebyshev representation \cite{canutoetal88}.

The number of mesh points required to capture the smallest scales in a
DNS is most of the time estimated following Kolmogorov's phenomenology
\cite{kolmogorov41a,kolmogorov41b}. Assuming that all the scales
smaller than the Kolmogorov scale $\eta$ are dissipated and cannot
contribute to the inertial range dynamics, it is usually established
that the number of mesh points $N$ required in a DNS of homogeneous
isotropic turbulence of dimension $L_b^3$ can be estimated with the
relation $N \approx (L_b/\eta)^{3} \sim Re_{L_b}^{9/4}$ where
$Re_{L_b}$ is the Reynolds number based on $L_b$ (which is
representative of the integral scale $L$) and the rms of the
fluctuating velocity $u^{\prime}$. This relation was first shown in
1959 by \cite{landau&lifshitz59} and is nowadays widely used in many
studies based on DNS. It is important to point out that this relation
is assuming local isotropy for the flow and an average dissipation
approximated with $\langle \varepsilon \rangle \approx C_{\varepsilon}
u^{\prime 3}/L_b$, with $C_{\varepsilon}$ defined as a constant
\cite{vassilicos15,ishiharaetal09}. It is also possible to estimate
the cost in terms of time steps $T/\Delta t$ with $T\approx
L_b/u^{\prime}$ corresponding to the timescale related to the
dimension $L_b$ of the cubic box. Assuming that $u^{\prime} \Delta t
\approx \Delta x$, we obtain $T/\Delta t \sim Re_{L_b}^{3/4}$ if
$\Delta x \sim \eta$ \cite{ishiharaetal09}.  These estimates can be
used to evaluate the computational power $W$ required to perform a
DNS. If we have $\eta \approx \Delta x$ then $W$ scales as
$(L_b/\Delta x)^3 (T/\Delta t)\approx Re_{L_b}^3$. This means that
doubling the Reynolds number requires nearly an order of magnitude
increase in computational effort.

In recent years several authors have been debating these relations and
whether they are accurate enough to evaluate high-order derivatives
and high order statistics. In \cite{donzisetal08}, the authors
investigated resolution effects and scaling in DNS of homogeneous
isotropic turbulence with a special attention to dissipation and
enstrophy, with resolutions of up to $\Delta x/\eta=0.25$. They
confirmed that the formula to evaluate the resolution of a DNS of
statistically steady forced periodic turbulence designed to resolve
the smallest scales of the flow within a constant multiple $a$ of
$\eta$ and with a computational box whose linear size $L_b$ is a
constant multiple $b$ of the largest scale of the flow, i.e. the
integral scale $L$, can be approximated as $N\approx 0.05 \frac{b}{a}
Re^{4.5}_{\lambda}$. This formula is very similar to the previous
formula, assuming that $Re_{\lambda} \sim Re_{L}^{0.5}$. They
showed that, in the context of statistically stationary homogeneous
isotropic forced turbulence, this standard resolution is adequate for
computing second-order quantities but is underestimating high-order
moments of velocity gradients. They demonstrated that the smallest
scale that needs to be resolved to capture high-order quantities (of
order $n$, with $n \rightarrow \infty$) is $\eta_{min}\approx L
Re^{-2}_{\lambda}$. In \cite{yakhot&sreenivasan05}, the authors state
that the computational power needed to perform a DNS of fully
developed homogeneous isotropic turbulence increases as $Re_{L}^4$
if ones want to study high-order quantities, and not as the
$Re_{L}^3$ expected from Kolmogorov’s theory.

In practice, the previous estimates can be a little tricky to use and
recent work suggests that they do not apply to a wide enough range of
turbulent flows. Indeed, it has recently been shown that
$C_{\epsilon}$ is not constant in a substantial region of spatially
evolving fully developed turbulent flows such as decaying
grid-generated turbulence and axisymmetric turbulent wakes
\cite{vassilicos15,laizet&vassilicos14,nedicetal13} and is also not
constant in unsteady periodic turbulence \cite{goto&vassilicos14}. In
all these cases $C_{\epsilon}$ is proportional to the ratio of a
global inlet/initial Reynolds number $Re_I$ to a local (in time or
space) Reynolds number $Re_{L}$ based on $u'$ and an integral
length-scale $L$. This has of course direct implications on DNS
resolution requirements as $C_{\epsilon}$ is taken to be constant in
the aforementioned resolution estimates. In the case of a DNS of
unsteady periodic turbulence, such as the one of
\cite{goto&vassilicos14} for example, this new scaling of
$C_{\epsilon}$ implies that $N \sim (L/\eta)^{3} \sim
C_{\epsilon}^{3/4} Re_{L}^{9/4} \sim Re_{I}^{3/4} Re_{L}^{3/2}$. In
the case of more realistic turbulent flows, and therefore more
complicated, than periodic turbulence, such as the turbulent flows
considered in this paper, the resolution estimates based on periodic
homogeneous isotropic turbulence are neither directly nor easily
applicable.

%In practice, the previous estimations of the cost for a DNS are
%relatively tricky to use and it is still unclear if they are accurate
%enough. Recently, it has been shown experimentally that for grid
%generated turbulence $C_{\epsilon}$ is not necessarily constant
%downstream of the grid
%\cite{seoud&vassilicos07,mazellier&vassilicos10,valente&vassilicos12,nagataeta%l13,discettietal13,hearst&lavoie14}.
%It has of course a direct implication for the evaluation of the
%resolution requirements for a DNS as in all the estimations
%$C_{\epsilon}$ is assumed to be constant. Furthermore, the previous
%tentatives to estimate the cost of a DNS are based on homogeneous
%isotropic turbulence which is a limited factor.

Even though it is now possible to reach relatively high Reynolds
numbers using DNS, only very limited comparisons with experimental
data have been documented in order to evaluate the quality of a DNS.
Comparisons between hot-wire anemometry and DNS were carried out for a
fully turbulent pipe flow at a Reynolds number $Re_c \approx 7000$
based on centreline velocity and pipe diameter
\cite{eggelsetal94}. The resolution of their DNS followed the rule
$\Delta \leq \pi \eta$ with the use of a uniform cylindrical mesh in
the three spatial direction.The agreement between numerical and
experimental results was excellent for the lower-order statistics
(mean flow and turbulence intensities) and reasonably good for the
higher-order statistics (skewness and flatness factors of the
normal-to-the-wall velocity fluctuations).

\cite{monty&chong09} performed single point hot-wire measurements in a
turbulent channel flow at $Re_{\tau} = 934$ and compared their data
with the DNS of [1] at the same friction velocity Reynolds number
$Re_{\tau}$. Results showed excellent agreement between the streamwise
velocity statistics of the two data sets. The spectra were also very
similar, however, throughout the logarithmic region the secondary peak
in energy was clearly reduced in the DNS results because of the DNS
box length, leading to the recommendation that longer box lengths
should be investigated.

In \cite{schlatteretal09} the authors performed the first direct
comparisons between DNS and wind tunnel hot-wire and oil-film
inerferometry measurements of a turbulent zero-pressure-gradient
boundary layer at Reynolds numbers up to $Re_{\theta} = 2500$. They
found excellent agreement in skin friction, mean velocity and
turbulent fluctuations. However, they did point out that such
comparisons can be made difficult by, for instance, the choice
tripping which can affect the onset of transition to turbulence.

Note, finally, that the resolution in all these DNS as well as in
other recent DNS of high Reynolds number periodic turbulence
\cite{yeungetal12}, turbulent boundary layers
\cite{silleroetal13,eiteletal14} and turbulent mixing layers
\cite{attili&bisetti12} are all following the rule $\eta < \Delta x <
  3\eta$.

Our goal in the present numerical work is to assess the quality and
resolution requirements of DNS of a spatially developing turbulent
flow generated by a single square grid \cite{zhouetal14} against
hot-wire measurements in a wind tunnel. We investigate how the
resolution affects the fluid motion and special attention is
given to the effects of the quality and reliability of the numerical
data on small-scale statistics. For this, we focus on the strain-rate and
rotation tensors (the symmetric and skew-symmetric parts of the
velocity gradient tensor respectively) through a detailed analysis of
Q-R diagrams \cite{tsinober09} at various locations downstream of the
single square grid on the centreline of the flow.

\begin{figure}
\centering
\includegraphics[angle=-0,width=0.459\textwidth]{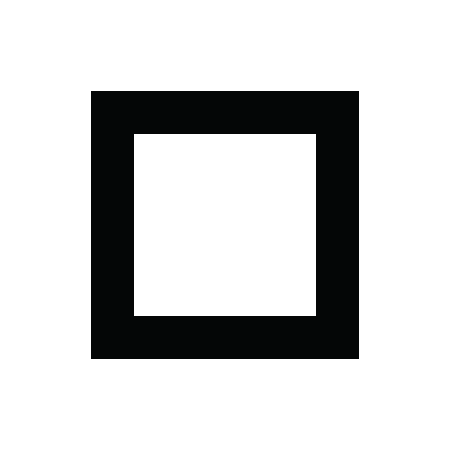}
\includegraphics[angle=-0,width=0.459\textwidth]{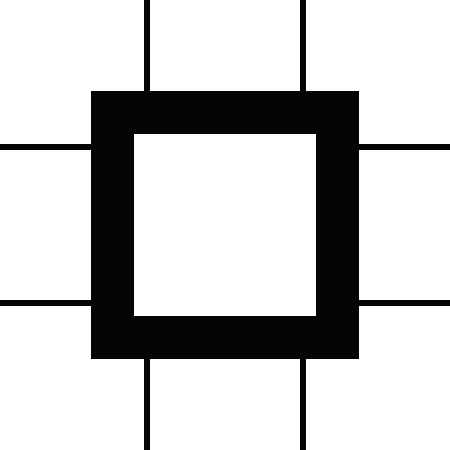}
\caption{Sketch of the single square grid used in the simulations
  (left) and in the experiments (right).}
\label{f:fig1}
\end{figure}

\section{Numerical set-up}

The single square grid represented in figure \ref{f:fig1} (left) is
defined using $L_0$ as the lateral length of each bar and $t_0$ as
their lateral thickness, with $L_0=5.30t_0$. The streamwise thickness
of the grid is $0.25t_0$. The computational domain $L_x \times L_y
\times L_z = 8L_0 \times 2L_0 \times 2L_0$ is discretized on a
Cartesian mesh using $n_x \times n_y \times n_z = 2881 \times 720
\times 720$ mesh nodes for the highly resolved $SSG$-$HR$ (Single
Square Grid - High Resolution) simulation, $n_x \times n_y \times n_z
= 1441 \times 360 \times 360$ mesh nodes for the low resolution
$SSG$-$LR$ (Single Square Grid - Low Resolution) simulation and $n_x
\times n_y \times n_z = 721 \times 180 \times 180$ mesh nodes for the
ultra low resolution $SSG$-$ULR$ (Single Square Grid - Ultra Low
Resolution) simulation. The coordinates $x$, $y$, and $z$ correspond
to the streamwise and the two cross-flow directions respectively. The
origin is placed at the centre of the single square grid, which is
located at a distance of $1.25L_0$ from the inlet of the computational
domain in order to avoid spurious interactions between the grid and
the inlet condition. The blockage ratio $\sigma$ for this single
square grid is $19\%$.  Inflow/outflow boundary conditions are used in
the streamwise direction while periodic boundary conditions are used
in the two lateral directions. The inflow condition is a uniform
profile $U_{\infty}$ free from any perturbations whereas the outflow
condition is a standard 1D convection equation.  For this numerical
work, after the evacuation of the initial condition, data are
collected over a period of $500,000$ time steps (with a time step of
$0.000139 L_0/U_{\infty}$).
% for the
%simulations $SSG$-$LR$ and $SSG$-$ULR$. For the simulation $SSG$-$HR$
%the data are collected over a period of $450,000$ to limit the
%computational effort.  
In terms of Reynolds number based on $L_0$, $Re_{L_O}=21,600$ for the
three simulations.

\begin{figure}
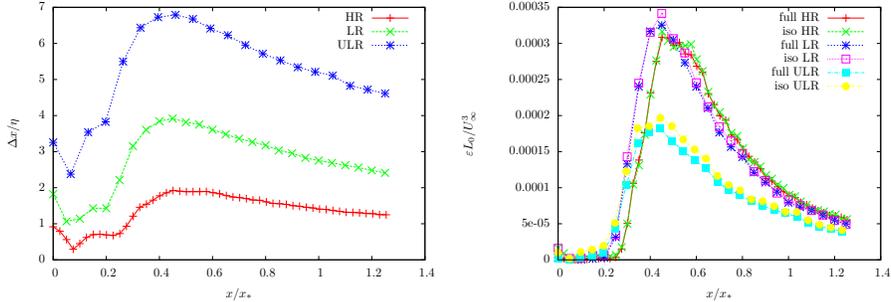

\centering
\resizebox{0.49\textwidth}{!}{\input{kolmo}}
\resizebox{0.49\textwidth}{!}{\input{diss_evol_ssg}}
\caption{Evolution of the spatial resolution with respect to the
  Kolmogorov microscale $\Delta x/ \eta$ (left), and of the normalised
  dissipation $\varepsilon L_0 / U^3_{\infty}$ (right) along the
  centreline where $x_* = L_0^2/t_0$, is the wake interaction
  length-scale \cite{gomesetal12}. For the single square grid case,
  the turbulence along the centreline reaches a maximum value at
  $x_{peak} \approx 0.5x_*$.}
\label{f:fig4b}
\end{figure}

In order to quantify the resolution with respect to the smallest
scales of the flow, we plot in figure \ref{f:fig4b} (left) the
streamwise evolution of $\Delta x/\eta$ along the centreline $y=z=0$,
where $\eta$ is the Kolmogorov microscale defined as
$(\nu^{3}/\epsilon_{full})^{1/4}$. The dissipation
$\varepsilon_{full}$ is evaluated using $\varepsilon_{full}=2\nu
\langle(\partial u^{\prime}_i/\partial x_j)^2 \rangle$. The full
dissipation and the dissipation $\varepsilon_{iso}=15\nu
\langle(\partial u_1^{\prime}/\partial x_1)^2 \rangle$, which is
obtained assuming the isotropy of the flow, are plotted in figure
\ref{f:fig4b} (right). It can be seen that the full dissipation and
the isotropic dissipation are virtually the same except maybe for the
simulation with the lowest resolution for which marginal differences
can be spotted. As expected, when the mesh is not fine enough to
capture the smallest scales of the flow, as for the $SSG$-$ULR$
simulation, the dissipation at the smallest scales cannot be taken
into account and is therefore underestimated by approximatively a
factor two. The $SSG$-$LR$ and $SSG$-$HR$ simulations are producing
similar levels for the dissipation.  For the simulation with the
finest mesh, $\Delta x/ \eta$ is always smaller than 2, whereas for
the coarsest mesh, $\Delta x/ \eta$ is always smaller than 7. As a
reference, in their recent very high Reynolds number Direct Numerical
simulations of wall bounded turbulence, the authors in
\cite{eiteletal14} have a comparable resolution with $\Delta x/ \eta
<2$. We are therefore expecting the mesh from the $SSG$-$HR$
simulation to be fine enough to take into account the smallest
features of the flow and a good comparison with experiments can be
expected.

\section{Numerical method}
The incompressible Navier-Stokes equations are solved using a recent
version of the high-order flow solver {\tt Incompact3d}\footnote{This
  open source code is now freely available at
  http://code.google.com/p/incompact3d/}, adapted to parallel
supercomputers using a powerful 2D domain decomposition strategy
\cite{laizet&li11}. This code is based on sixth-order compact finite
difference schemes for the spatial differentiation and an explicit
third order Adams-Bashforth scheme for the time integration. To treat
the incompressibility condition, a fractional step method requires
solving a Poisson equation. This equation is fully solved in spectral
space, via the use of relevant 3D Fast Fourier Transforms. The
pressure mesh is staggered from the velocity mesh by half a mesh, to
avoid spurious pressure oscillations. With the help of the concept of
modified wave number, the divergence-free condition is ensured up to
machine accuracy. The modelling of the grid is performed using an
Immersed Boundary Method (IBM) based on direct forcing approach that
ensures the no-slip boundary condition at the obstacle walls. The idea
is to force the velocity to zero at the wall of the single square
grid, as our particular Cartesian mesh does conform with the
geometries of the grid. It mimics the effects of a solid surface on
the fluid with an extra forcing in the Navier-Stokes equations. Full
details about the code can be found in \cite{laizet&lamballais09}.

\section{Experimental set-up}
In order to provide data for comparison with the simulations,
experiments were conducted with a very similar grid, as shown in
figure \ref{f:fig1} (right), in a blow down wind tunnel, whose working
section has a square cross-section of 0.4572m $\times$ 0.4572m (18"
$\times$ 18") and a working length of 3.5m, with the turbulence
generating grids placed at the start of the test-section. The
background turbulence level is 0.1\%. A grid is installed at the
entrance of the diffuser to maintain a slight over-pressure in the
test section. The inlet velocity $U_{\infty}$ is controlled using the
static pressure difference across the 8:1 contraction, the temperature
taken near the diffuser and the atmospheric pressure from a pressure
gauge, all of which were measured using a Furness Controls
micromanometer FCO510.

Measurements of the velocity signal on the centreline of the flow were
taken using a DANTEC 55P01 hot-wire ($5\mu m$ in diameter with a
sensing length of 1.25mm), driven by a DANTEC Streamline anemometer
with an in-build signal conditioner running in constant-temperature
mode (CTA). Data was sampled using a 16-bit National Instruments
NI-6229(USB) data acquisition card for $300sec$ at a sampling
frequency of 100kHz, with the analogue low-pass filter on the
Streamline set to 30kHz. Each data set was then digitally filtered,
using a fourth-order Butterworth filter to eliminate high frequency
noise, at a frequency of $f_c \ge 1.5 f_{\eta}$, where $f_{\eta} =
\langle u \rangle / 2\pi \eta$, where $\langle u \rangle$ is the local
mean streamwise velocity. The blockage ratio for the grid in the
experiments is slightly larger than the one in the simulations with a
value of $21\%$, because of the addition of the small bars to hold the
single square grid in the wind tunnel. In terms of Reynolds numbers
based on $L_0=228.6mm$, $Re_{L_O}=36450$, $72,900$ and $145,900$,
corresponding to an inflow velocity $U_{\infty}$ of $2.5m/s$, $5m/s$
and $10m/s$.

Note that the Reynolds number $Re_{L_{0}}$ in the three simulations is
about 1.7 times smaller than the smallest Reynolds number of the
experiments as it was not possible to reduce the speed of the wind
tunnel below $2.5m/s$.  Because of computational constraints for the
simulation with the highest resolution, the collection time $T =
16sec$ for the simulations, corresponding to $500,000$ time-steps, is
much smaller than in the experiments where it is $T = 300sec$. One
substantial difference between the experimental and numerical set-ups
is in the boundary conditions, walls in the wind tunnel as opposed to
periodic boundary conditions in the simulations. However, because of
the low blockage ratio of the single square grid, we do not expect a
significant impact of the walls and/or the boundary conditions on the
centreline of the grid where we evaluate the quality of the
simulations.

\begin{figure}[h]
\centering
\includegraphics[trim=0cm 0cm 0cm 15cm,width=0.8\textwidth,clip=]{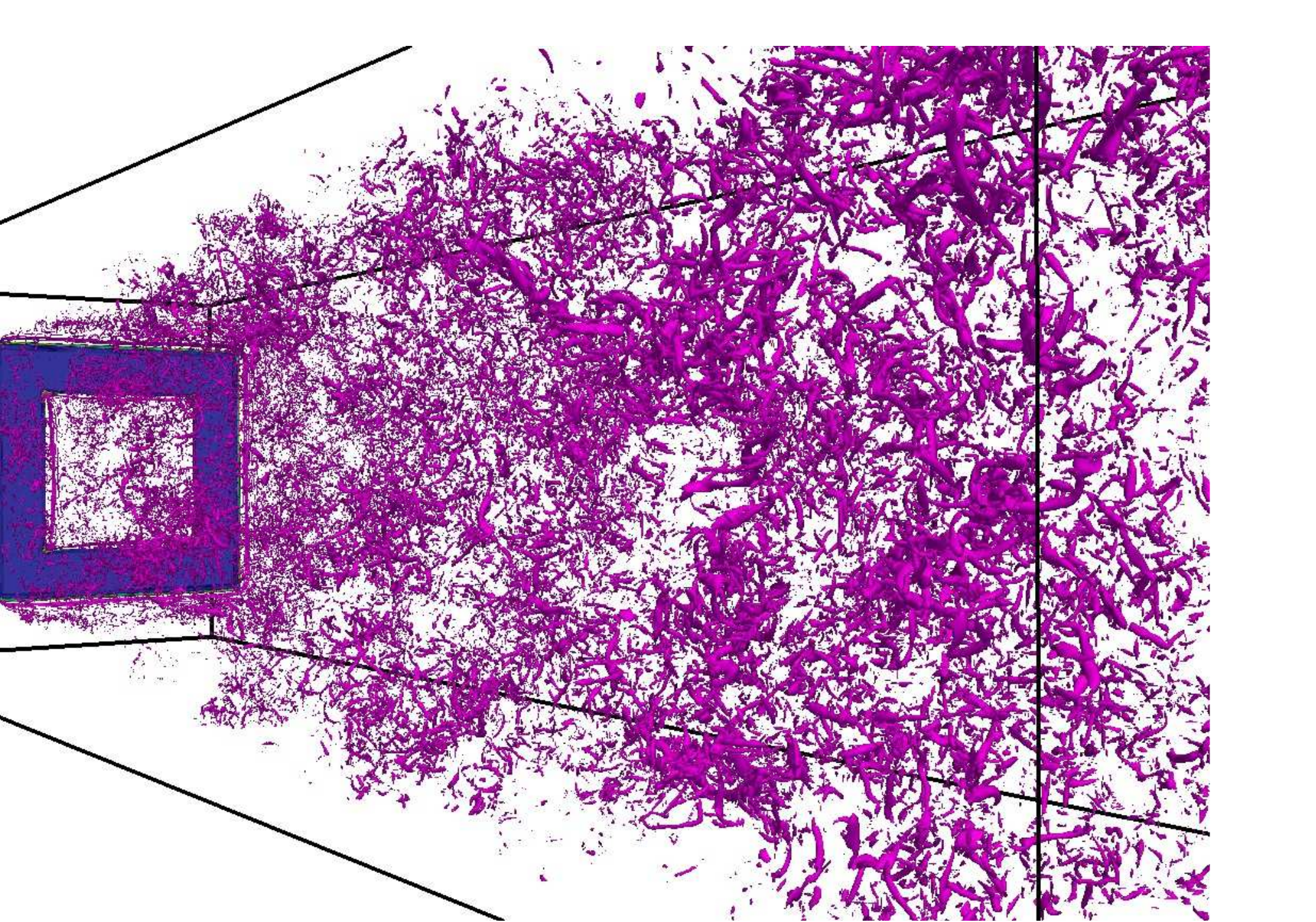}
\caption{Turbulent flows generated by the single square grid for the
  $SSG$-$HR$ simulation; 3D isosurfaces equal to 0.5 of the absolute
  value of the enstrophy vector normalised by its maximum over the
  ($y-z$) plane at the $x$-position considered.}
\label{f:3dvisu}
\end{figure}

\section{Comparison with experiments}

An illustration of the flow obtained downstream of the single square
grid is given in figure \ref{f:3dvisu}, where enstrophy isosurfaces
are plotted. These isosurfaces are showing the enstrophy normalised by
its maximum over the ($y-z$) plane at the $x$-position
considered. The four same-size wakes generated by the four bars of the
grid interact and mix together to give rise to a fully turbulent
flow. 

\begin{figure}
\centering
\includegraphics[angle=-0,width=0.48\textwidth]{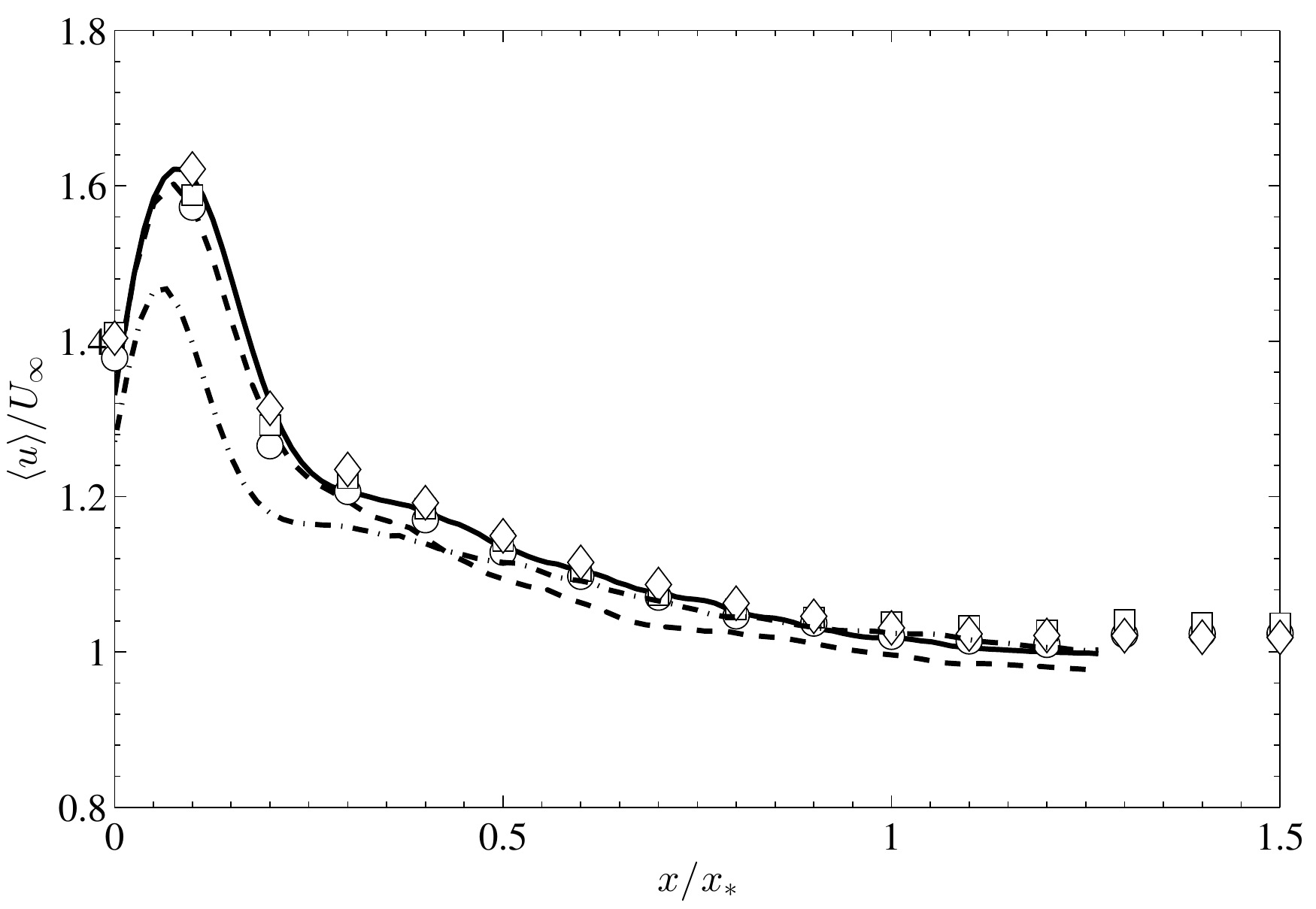}
\includegraphics[angle=-0,width=0.49\textwidth]{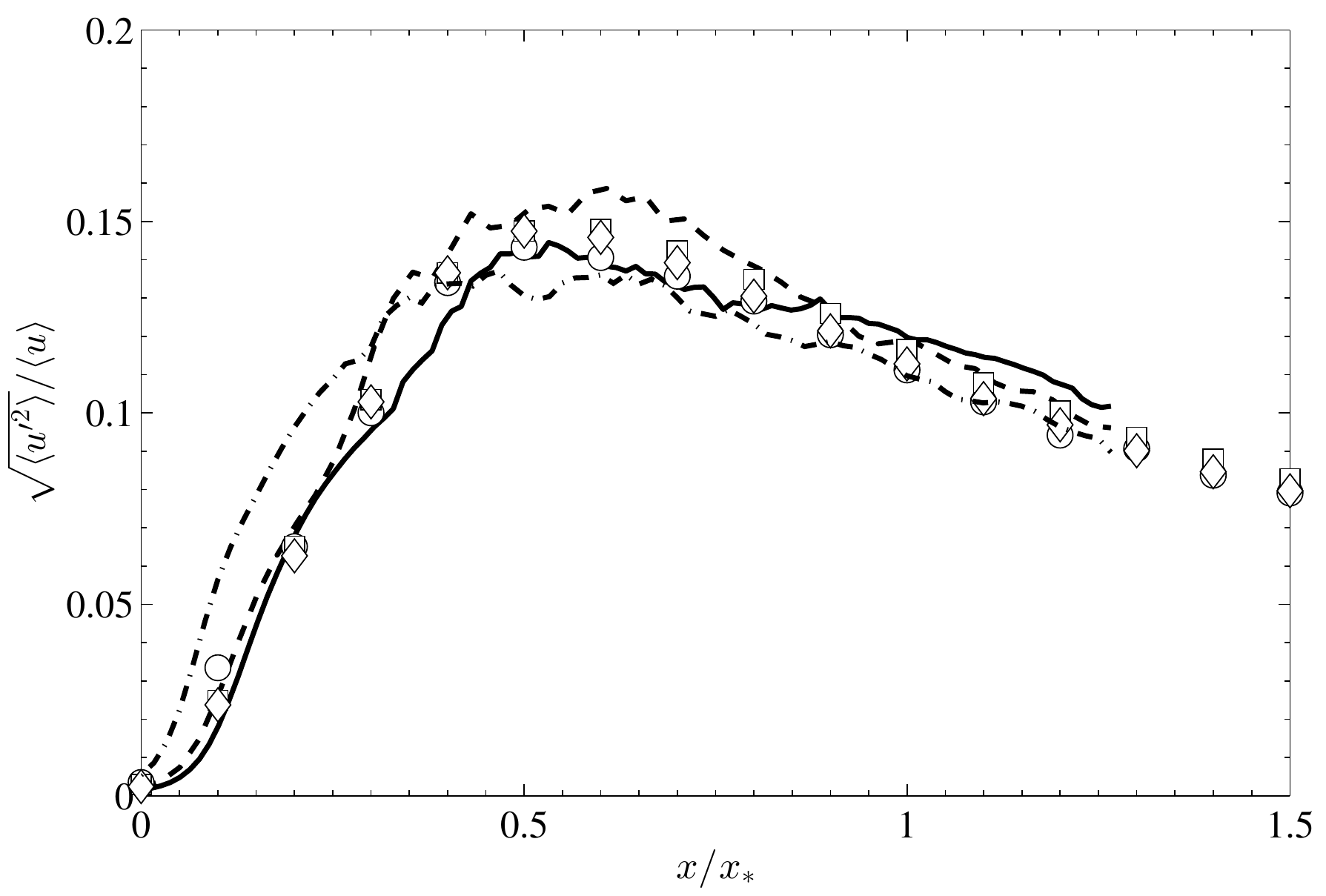}
\caption{Evolution of the local mean streamwise velocity $\langle u
  \rangle$ normalised with $U_{\infty}$ (left) and of the streamwise
  turbulence intensity $\sqrt{\langle u \rangle^{\prime 2}}$
  normalised with the local mean streamwise velocity $\langle u
  \rangle$ (right) along the centreline. The continuous line
  corresponds to $SSG$-$HR$, the dashed line to $SSG$-$LR$ and the
  dot-dashed line to $SSG$-$ULR$. The symbols correspond to the
  experiments with $\Diamond$$=10m/s$, $\Box$$=5m/s$,
  {\Large{$\circ$}}$=2.5m/s$.}
\label{f:fig3}
\end{figure}

We first compare, along the centreline of the flow, the streamwise
evolutions of the local mean streamwise velocity $\langle u \rangle$
and of the streamwise turbulence intensity $\sqrt{\langle u
  \rangle^{\prime 2}}$. The results, presented in figure \ref{f:fig3},
are normalised with $x_* = L_0^2/t_0$ which is the wake interaction
length-scale \cite{gomesetal12}. For the single square grid case, the
turbulence along the centreline reaches a maximum value at $x_{peak}
\approx 0.5x_*$ both for experimental and numerical data. For the
experimental data, it can be seen that there is a maximum value of
about 1.6 for $\langle u \rangle/U_{\infty}$ located at $x=0.1x_*$,
followed by a fast decay up to $x=0.3x_*$, and then a slow decay up to
$x=x_*$. After that point, the effect of the boundary layers at the
wall of the wind tunnel can eventually be seen for the experimental
data with a very slow increase of $\langle u \rangle/U_{\infty}$. The
$SSG$-$HR$ and $SSG$-$LR$ simulations are in very good agreement with
the experiments both quantitatively and qualitatively, with for
instance the correct prediction of the maximum value 1.62 at
$x=0.1x_*$. The data for the $SSG$-$HR$ simulation and the experiments
are even on top of each other up to $x=x_*$. The $SSG$-$ULR$
simulation is under predicting the streamwise evolution of $\langle u
\rangle/U_{\infty}$. For instance, the maximum value for $\langle u
\rangle$ is under estimated by about $10\%$ with a value of only 1.467
at $x=0.075x_*$. Concerning the evolution of $\sqrt{\langle u
  \rangle^{\prime 2}}$, only the $SSG$-$HR$ simulation is able to
predict correctly the location of the peak and its intensity. An
important result here is the over estimation by SSG-ULR of
$\sqrt{\langle u \rangle^{\prime 2}}$ in the production region before
the peak of turbulence. This could be attributed to a pile-up of
energy due to the lack of dissipation in the simulation, with the
creation of numerical spurious oscillations just downstream of the
single square grid where the resolution for the smallest scales is not
good enough. After the peak, the three simulations are in relative
good agreement with the experimental data, even in those cases where
the location of the peak of turbulence was not predicted properly.
%\begin{figure}
%\centering
%\includegraphics[angle=-0,width=0.95\textwidth]{2D_map_umean.ps}
%\caption{2D map in the $(x-y)$ plane for $z=0$ for $\langle u \rangle$
%  normalised with $U_{\infty}$. From top to bottom: $SSG$-$HR$,
%  $SSG$-$LR$ and $SSG$-$ULR$.}
%\label{f:fig31}
%\end{figure}

\begin{figure}
\centering
\includegraphics[trim=0cm 0cm 0cm 0cm,width=0.95\textwidth,clip=]{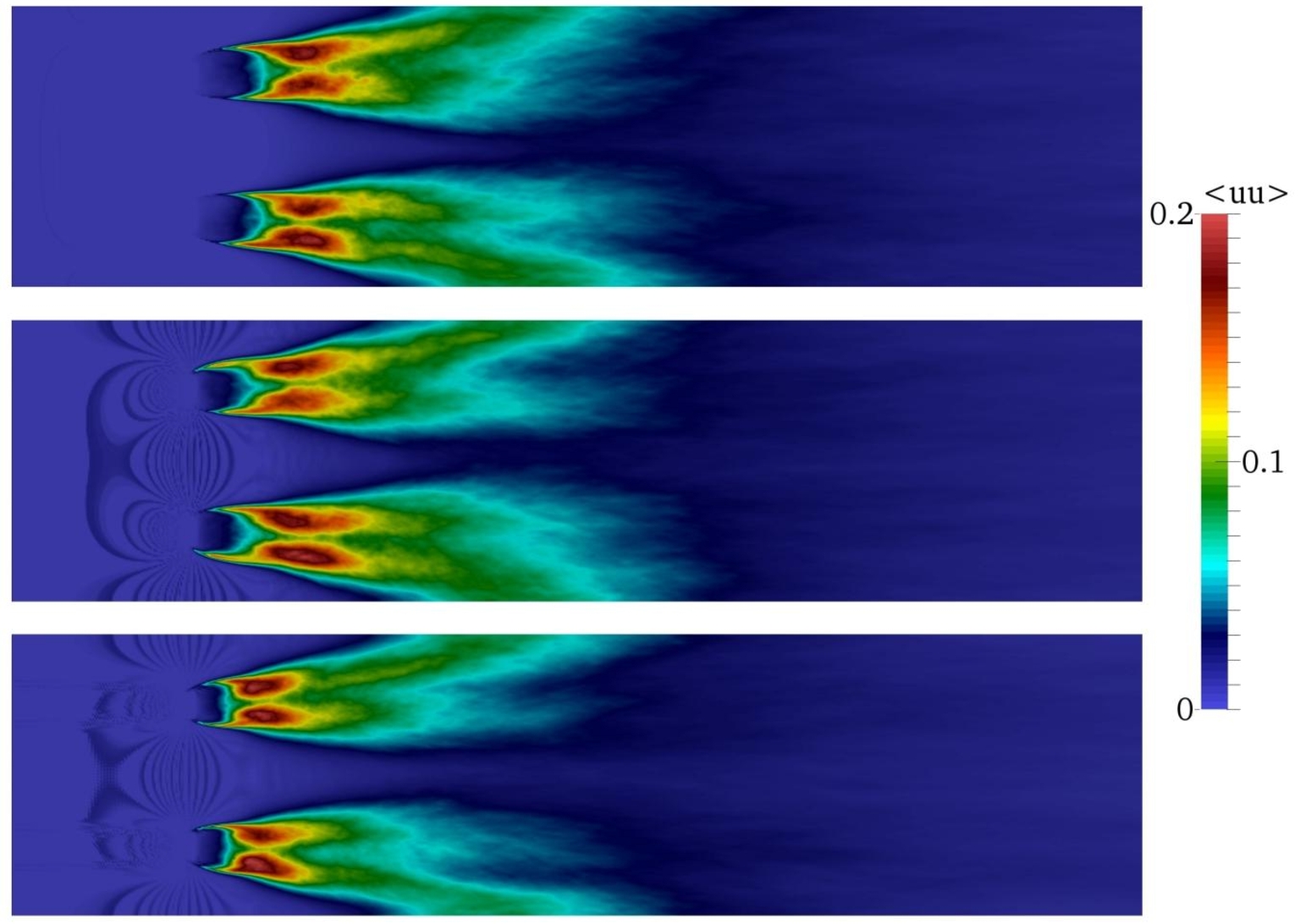}
\caption{Contour map of $<u^{\prime 2}>/U_{\infty}^2$ in the $xy$
  plane at z=0. From top to bottom: $SSG$-$HR$, $SSG$-$LR$ and
  $SSG$-$ULR$.}
\label{f:fig32}
\end{figure}

The effect of the resolution can clearly be seen in figure
\ref{f:fig32} where a 2D map of $\langle u^{\prime 2}
\rangle/U^2_{\infty}$ is plotted in the $(x-y)$ plane for $z=0$. In
front of the grid, spurious oscillations appear when the resolution is
not good enough. It is the signature of the pile-up of energy at the
small scales. As already observed on the centreline, these
oscillations seem to have to have a relatively low impact on the
dynamic of the flow downstream of the grid.

\begin{figure}
\centering
\includegraphics[angle=-0,width=0.49\textwidth]{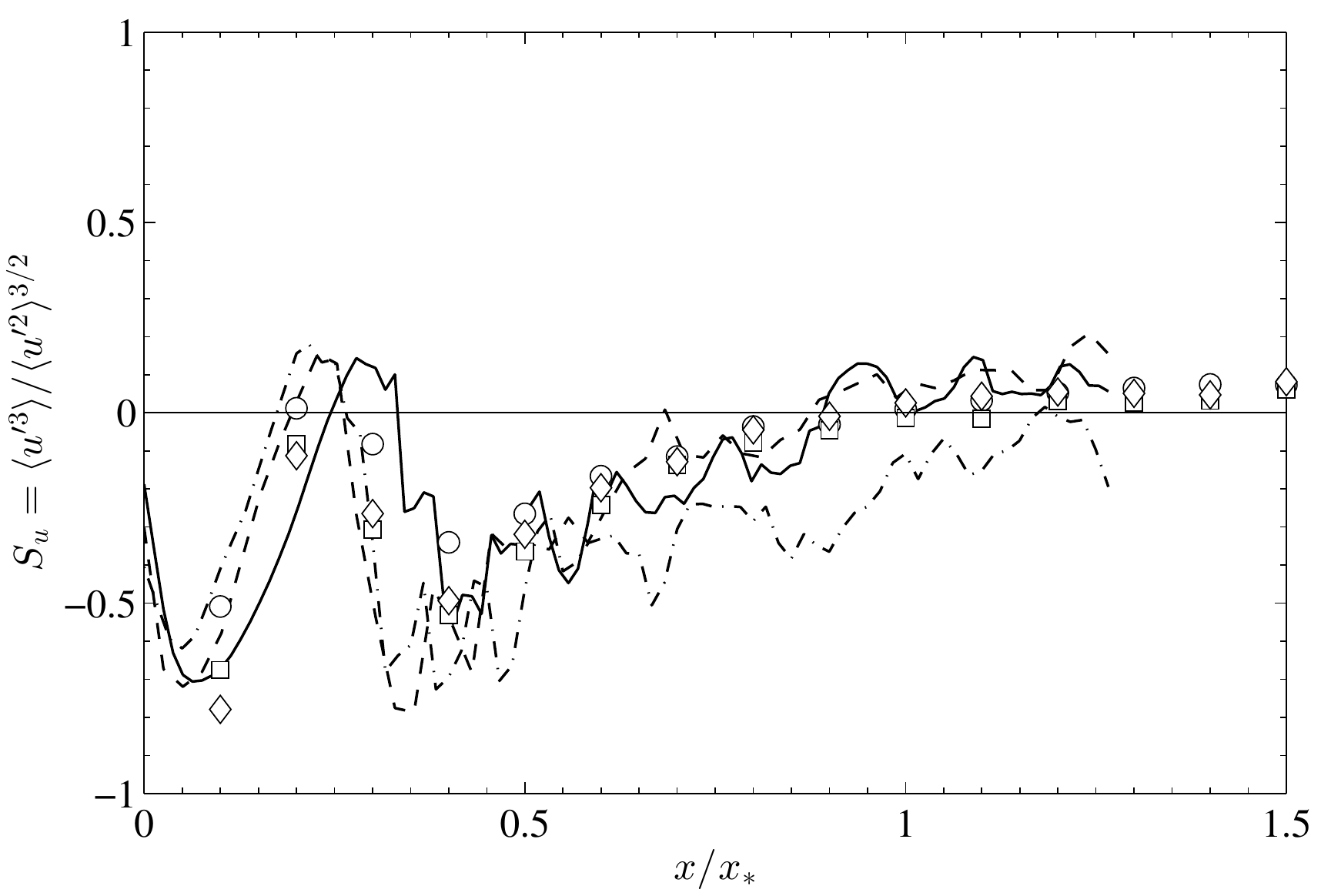}
\includegraphics[angle=-0,width=0.48\textwidth]{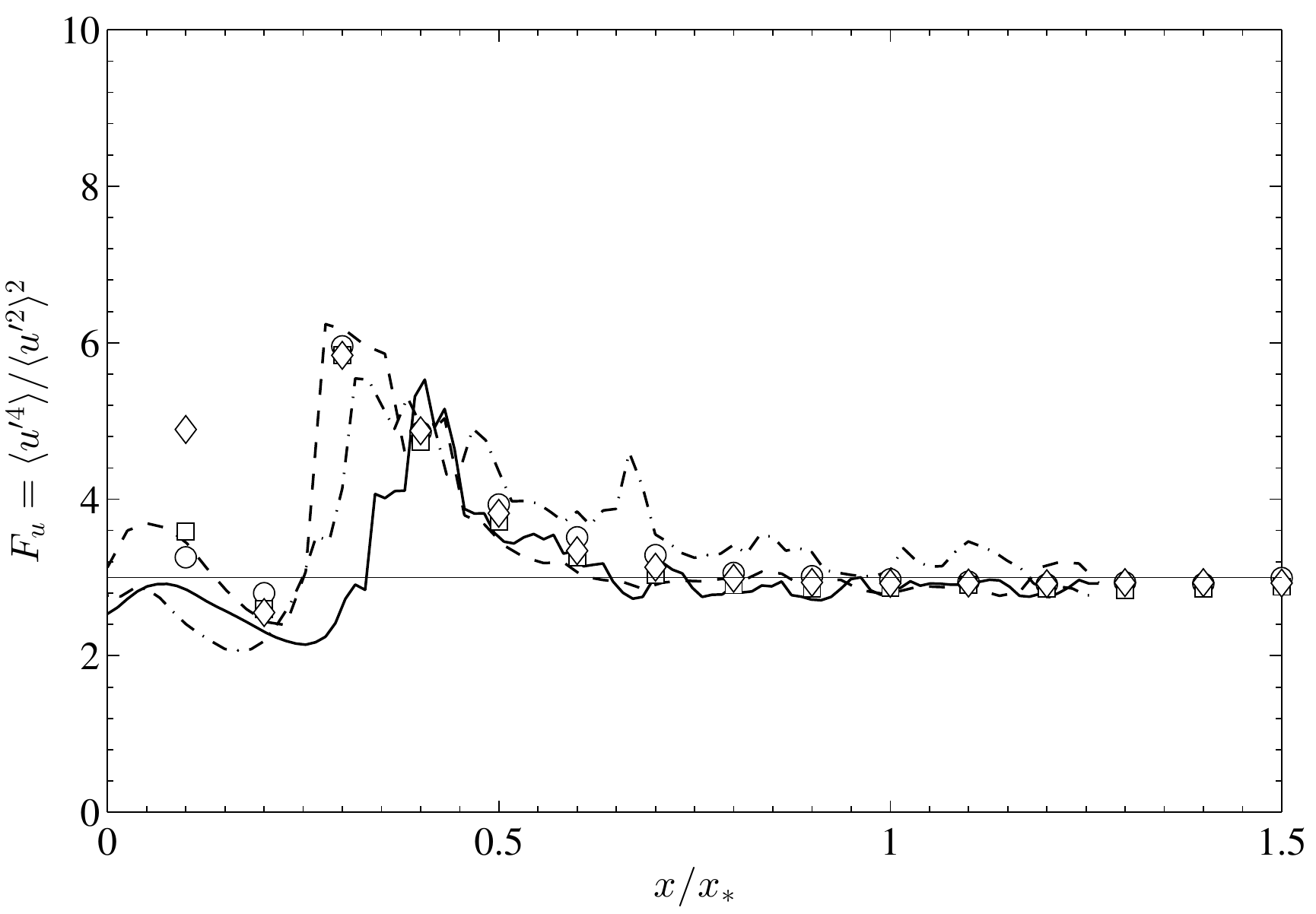}
\caption{Evolution of the skewness (left) and the flatness (right) of
  the fluctuating streamwise velocity. The continuous line corresponds
  to $SSG$-$HR$, the dashed line to $SSG$-$LR$ and the dot-dashed line
  to $SSG$-$ULR$. The symbols correspond to the experiments with
  $\Diamond$$=10m/s$, $\Box$$=5m/s$, {\Large{$\circ$}}$=2.5m/s$.}
\label{f:fig4}
\end{figure}

Figure \ref{f:fig4} shows the streamwise evolution of the skewness
$S_u$ and of the flatness $F_u$ of the streamwise turbulence intensity
where
\begin{equation}
S_u=\frac{\langle u^{\prime 3}\rangle}{\langle u^{\prime
    2}\rangle^{3/2}},~~~~~~~~~F_u=\frac{\langle u^{\prime
    4}\rangle}{\langle u^{\prime 2}\rangle^{2}}
\end{equation}

It is clear that the numerical data are not converged enough to get a
smooth profile for the streamwise evolution.  However, the numerical
data are in good quantitative agreement with the experiments as they
are following the same trend. In the production region, the values
obtained for the skewness and the flatness suggest that the
distribution of the velocity is highly non-Gaussian both for the
experiments and for the simulations. This is related to the presence
of strong events in the flow, as reported previously by
\cite{mazellier&vassilicos10,zhouetal14}, in the production region of
grid-generated turbulence. After the peak of turbulence located at
$x=0.5x_*$, the skewness is converging to zero and the flatness is
converging to 3, corresponding to a Gaussian distribution for the
streamwise turbulence intensity.

\begin{figure}
\centering
\includegraphics[angle=-0,width=0.49\textwidth]{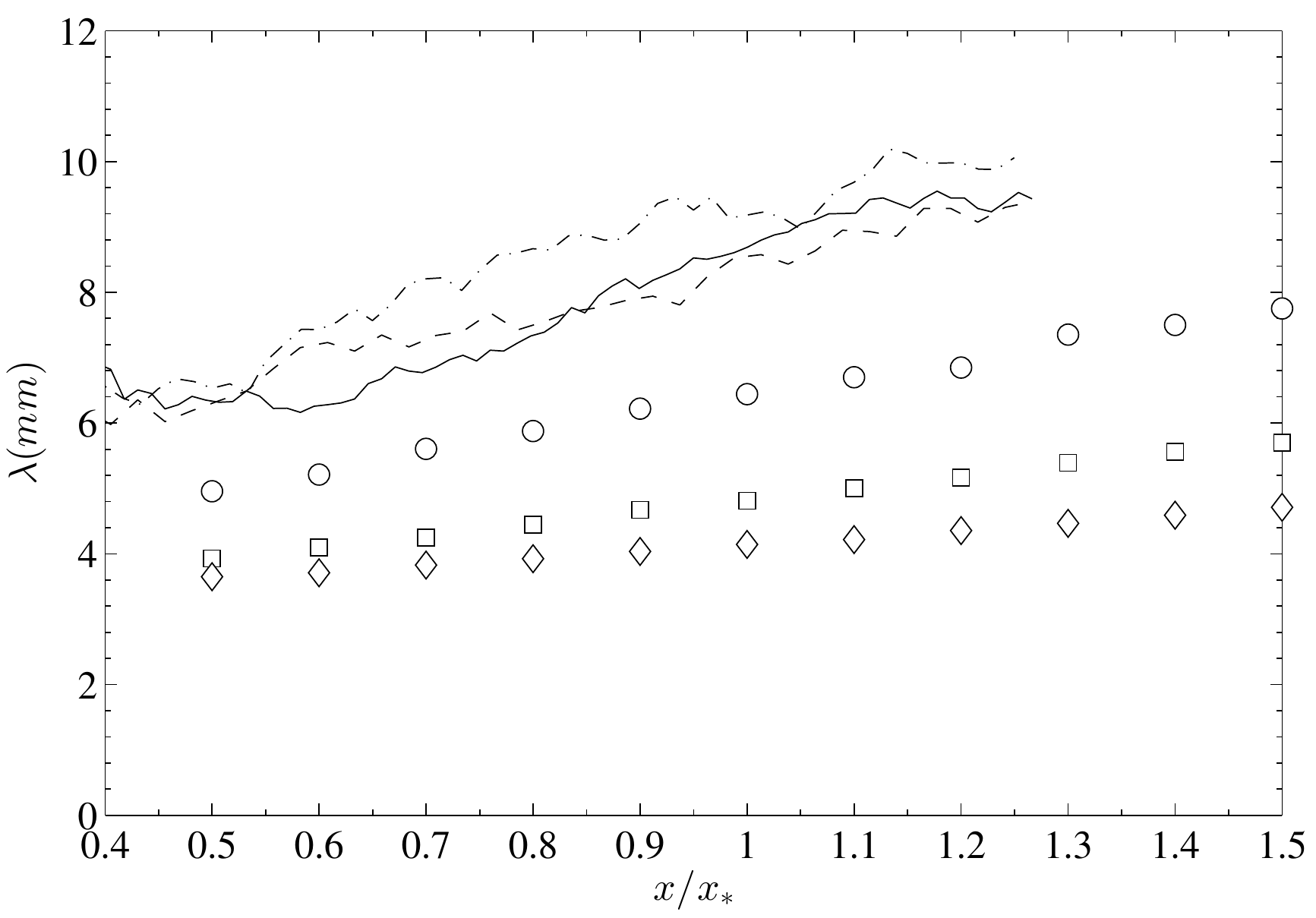}
\includegraphics[angle=-0,width=0.49\textwidth]{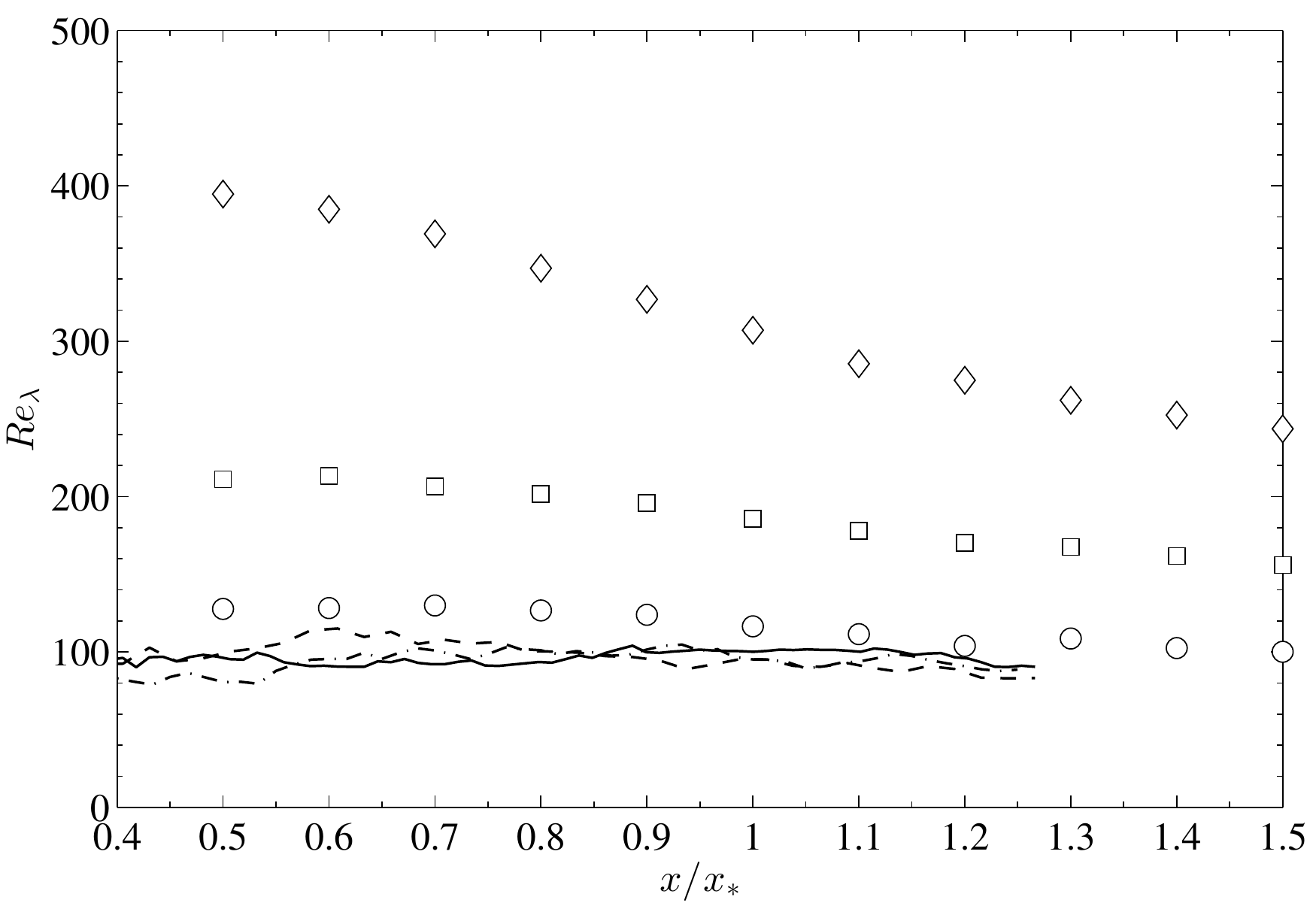}
\caption{Evolution of the Taylor micro-scale $\lambda$ (left) and of
  the associated local Reynolds number (right). The continuous line
  corresponds to $SSG$-$HR$, the dashed line to $SSG$-$LR$ and the
  dot-dashed line to $SSG$-$ULR$. The symbols correspond to the
  experiments with $\Diamond$$=10m/s$, $\Box$$=5m/s$,
  {\Large{$\circ$}}$=2.5m/s$.}
\label{f:fig41}
\end{figure}

Figure \ref{f:fig41} shows the streamwise evolution of the Taylor
micro-scale $\lambda$ and of the associated local Reynolds
number, where
$$ 
\lambda \approx \sqrt{\frac{\langle u^{\prime 2} \rangle}{\left \langle \left(
  \frac{1}{\langle u \rangle}\frac{\partial u'}{\partial t}\right)^2
  \right \rangle} }
$$ 
As expected, $\lambda$ is slowly increasing after the peak of
turbulence (located at $\approx 0.5x_*$) when moving downstream of the
grid. When the inflow velocity is increased, $\lambda$ is reduced. The
values obtained between $4mm$ and $10mm$ are consistent with previous
values obtained experimentally for a fractal square grid by
\cite{valente&vassilicos11} where values between $4mm$ and $6mm$ were
reported for an inflow velocity of $15m/s$. A notable result is that
for the DNS data and for the experimental data with the lowest
velocity, $Re_{\lambda}$ remains constant after the peak of turbulence
(located at $\approx 0.5x_*$).  It is a rather surprising result which
could be attributed to the low inflow velocity of the flow but which
is in good agreement with the numerical results of
\cite{zhouetal14}. In \cite{valente&vassilicos12} it was shown
experimentally that for a very similar grid, $Re_{\lambda}$ was
decreasing after the peak of turbulence, which is consistent with the
current experimental data. The present results are therefore
suggesting that the new dissipation law \cite{vassilicos15}, for which
$C_{\epsilon}$ is not constant, is valid above a certain value for
$Re_{\lambda}$, value which is grid-dependant.

\begin{figure}
\centering
\includegraphics[angle=-0,width=0.49\textwidth]{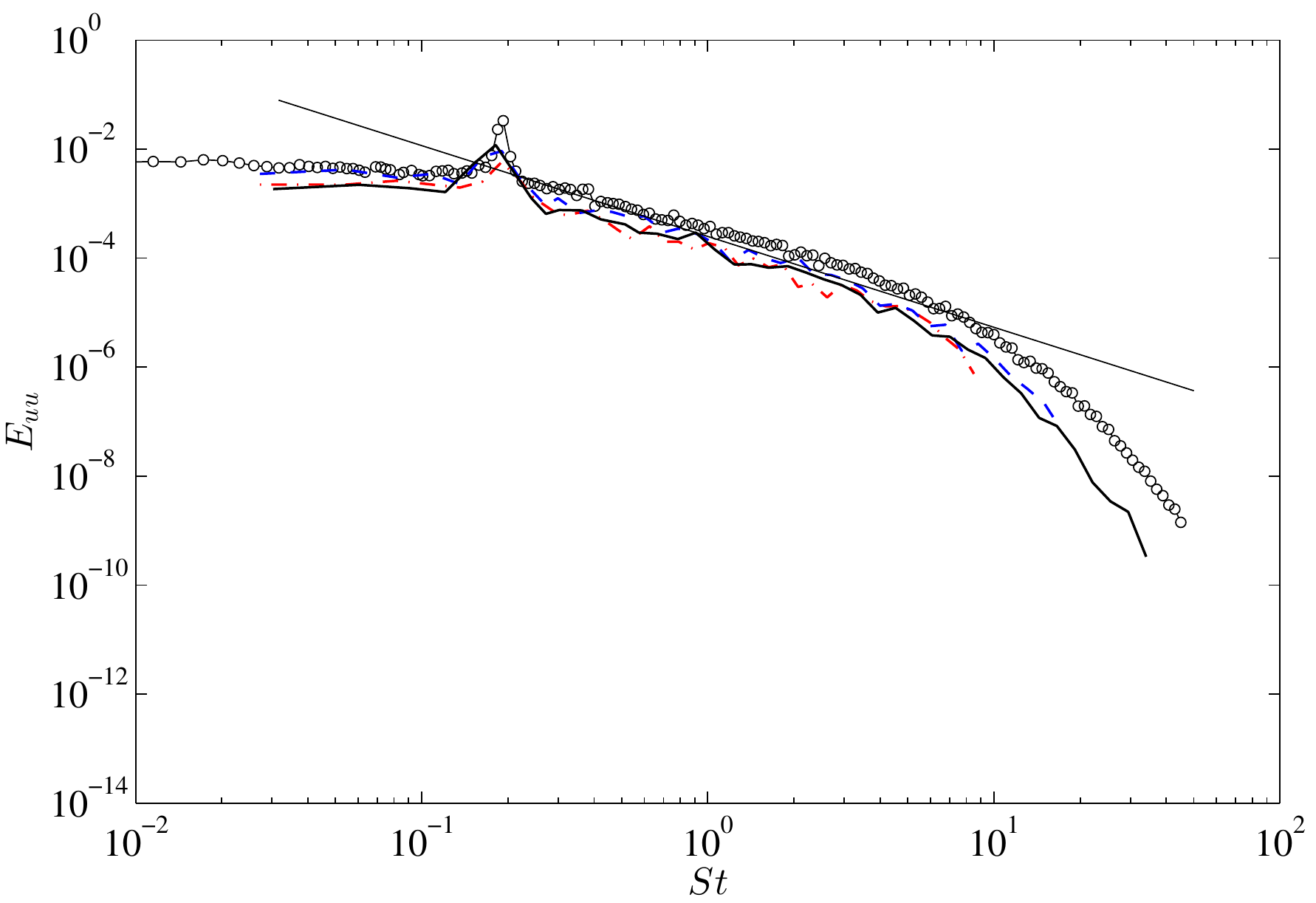}
\includegraphics[angle=-0,width=0.49\textwidth]{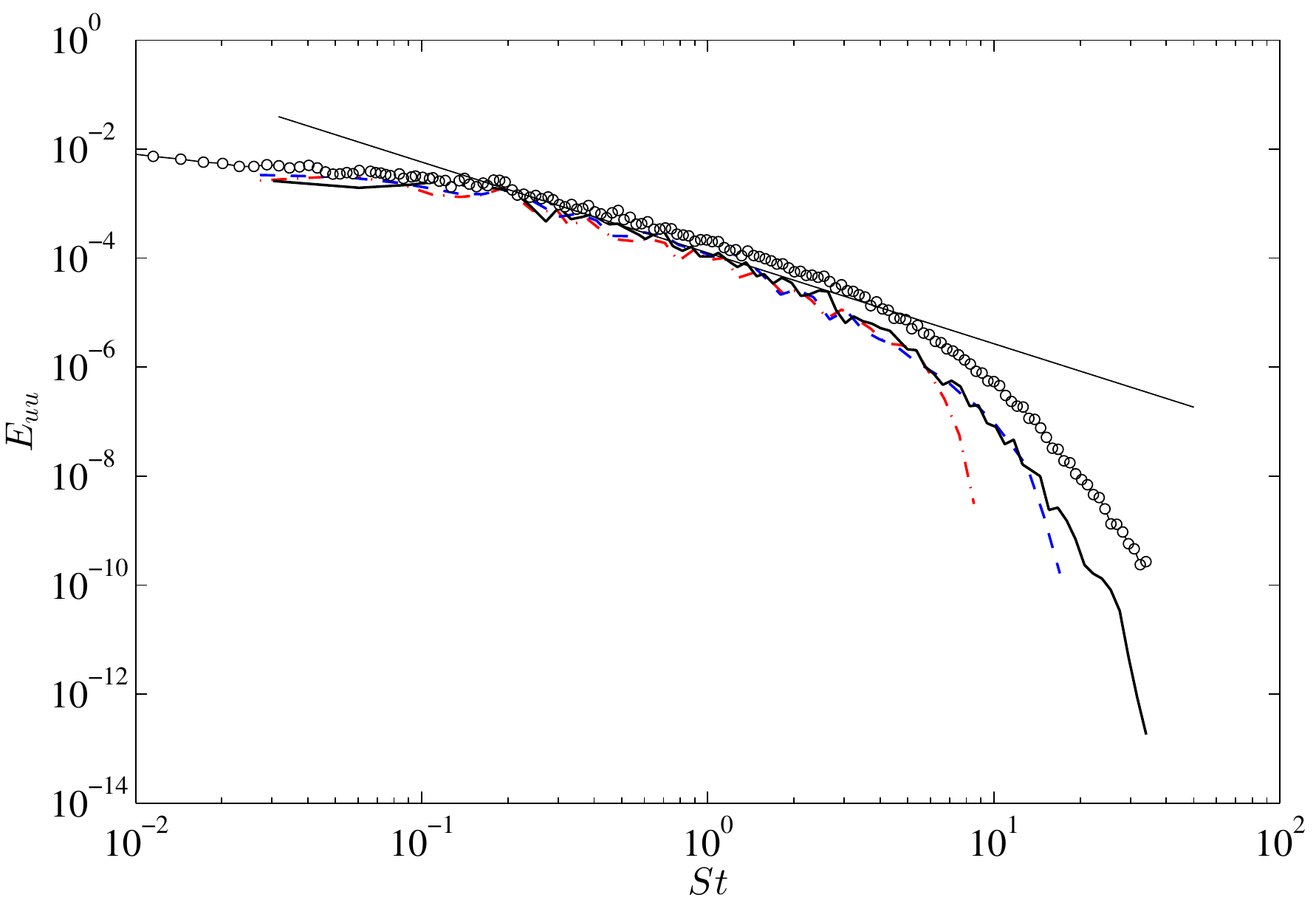}
\caption{Energy spectra of the streamwise fluctuating velocity
  component at $0.4x_*$ (left) and at $x_*$ (right). The continuous
  line corresponds to $SSG$-$HR$, the dashed line to $SSG$-$LR$ and
  the dot-dashed line to $SSG$-$ULR$. The symbols correspond to the
  experiments with $U_{\infty}=2.5m/s$.}
\label{f:fig5}
\end{figure}

In order to investigate a bit further the quality of the simulations,
the energy spectra obtained at $0.4x_*$ and at $x_*$ for the
streamwise turbulence intensity $\sqrt{u^{\prime 2}}$ are presented in
figure \ref{f:fig5}. These energy spectra, obtained in the frequency
domain on the centreline of the flow are estimated using the
periodogram technique \cite{pressetal92}. Data are collected in time
for the simulations using virtual probes in a similar fashion to the
experiments. The time signal is then divided in several sequences with
an overlap of $50\%$ with the use of a Hanning window. The cut-off
frequency for the simulations is $f_c=U_{\infty}/2\Delta x$,
corresponding to the smallest frequency that the mesh can see and for
the experiments we have $f_c=1.5f_{\eta}$, where $f_{\eta} = \langle u
\rangle / 2\pi \eta$.  The energy spectra plots obtained at $0.4x_*$
in the production region are showing a very good agreement between the
experiments and the simulations with the correct prediction of the
Strouhal, corresponding to the frequencies of the large scale vortices
generated by the single square grid. As expected, the levels of energy
are slightly larger in the experiment, as the inflow velocity and
therefore the global Reynolds number are higher.  Interestingly
enough, the energy spectra for the three simulations in the production
region are following the same trend and it is difficult to observe any
resolution effect near the cut-off frequency of each
simulations. However, in the decay region after the peak of
turbulence, there is a clear drop-off of the energy spectra near the
cut-off frequency for the three simulations, as seen in figure
\ref{f:fig5} for $x=x_*$. Note also that the cut-off frequency of the
experiments is the same as the one for the high-resolution simulation,
suggesting that this simulation is able to capture the smallest scales
of the flow. From a physical point of view, both the experiments and
the simulations exhibit -5/3 frequency spectra for at least one decade
of frequencies.

\section{Effect of the resolution on the turbulence}

Following our previous work with fractal square grids
\cite{laizetetal13} and the recent work of \cite{zhouetal14} with a
single square grid, it is possible to obtain some information about
the resolution effects on vorticity and strain rate statistics using
the $Q$-$R$ diagram \cite{tsinober09}.

The velocity gradient tensor $A_{ij}=\partial u_i/\partial x_j$ can be
decomposed in a symmetric part $S_{ij}=(\partial u_i/\partial
x_j+\partial u_j/\partial x_i)/2$ and an anti-symmetric part
$W_{ij}=(\partial u_i/\partial x_j-\partial u_j/\partial
x_i)/2$. $S_{ij}$ is defined as the strain rate tensor and $W_{ij}$ as
the rotation rate tensor. Eigenvalues of $A_{ij}$ satisfy the
following characteristic equation
\begin{equation}
\lambda^3+P\lambda^2+R\lambda+R=0,
\end{equation}
with
\begin{equation}
P=-A_{ii},
\end{equation}
\begin{equation}
Q=-\frac{1}{2}A_{ij}A_{ji},
\end{equation}
\begin{equation}
R=-\frac{1}{3}A_{ij}A_{jk}A_{ki}.
\end{equation}

When the flow is incompressible, then $P=0$. Furthermore, one can
decompose $Q$ and $R$ as
\begin{equation}
Q=\frac{1}{4}(\omega_i\omega_i-2S_{ij}S_{ij})=Q_w + Q_s,
\end{equation}
with $Q_{w}= \frac{1}{4}\omega_i\omega_i$,
$Q_{s}=-\frac{1}{2}S_{ij}S_{ij}$ and
\begin{equation}
R=-\frac{1}{3}(S_{ij}S_{jk}S_{ki}+\frac{3}{4}\omega_i\omega_jS_{ij})=R_w + R_s,
\end{equation}
with $R_s=-\frac{1}{3}S_{ij}S_{jk}S_{ki}$,
$R_w=-\frac{1}{4}\omega_i\omega_jS_{ij}$ and
$\omega_i=\varepsilon_{ijk}\partial u_j/\partial x_k$,
$\varepsilon_{ijk}$ being the Levi-Civita symbol.

The $Q$-$R$ diagram has a tear drop shape in many turbulent flows
(turbulent boundary layers, mixing layers, grid turbulence, jet
turbulence) and according to \cite{tsinober09}, this tear drop shape
may be one of the qualitatively universal features of turbulent flows.
Therefore, it is a good indicator to assess the quality of our
simulations and check how the lack of resolution can affect the
$Q$-$R$ diagram.

\begin{figure}
\centering 
\resizebox{0.6\textwidth}{!}{\input{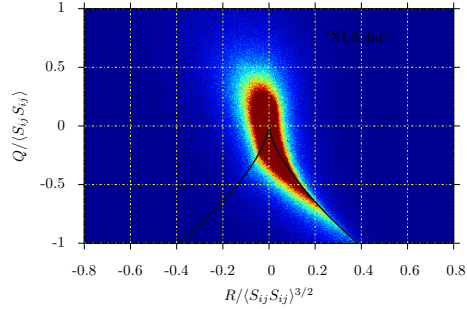}}
\caption{Joint probability density function of $Q$ and $R$ for
  periodic, statistically stationary turbulence from a DNS courtesy of
  Dr. R. Onishi, see \cite{onishietal11} for details. The dark red
  colour corresponds to isovalues greater than 0.025. Note that these
  statistics were obtained over all space at single snapshot in time.}
\label{f:fig7}
\end{figure}

As a reference, we are using the numerical data of a DNS of
periodic statistically stationary turbulence \cite{onishietal11}. The
$Q$-$R$ diagram obtained from a single time shot is presented in
figure \ref{f:fig7}. Note that $Q$ is normalised with $\langle
S_{ij}S_{ij} \rangle$ and $R$ by $\langle S_{ij}S_{ij}
\rangle^{3/2}$. As expected we can observed the tear drop shape when
$Q<0$ and $R>0$.

\begin{figure}
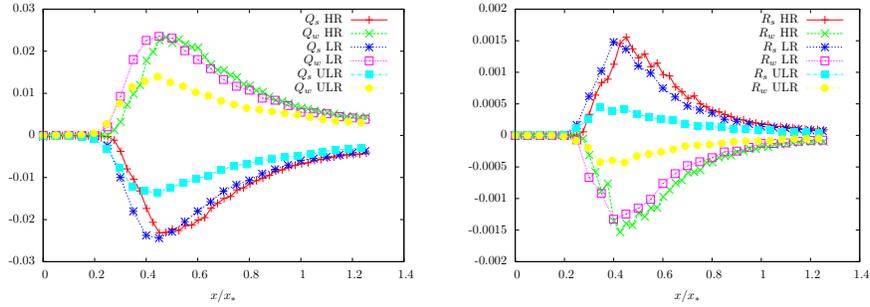

\centering
\resizebox{0.49\textwidth}{!}{\input{Q_evol_ssg}}
\resizebox{0.49\textwidth}{!}{\input{R_evol_ssg}}
\caption{Streamwise evolution of $\langle Q_{w}\rangle$, $\langle
  Q_{s}\rangle$, $\langle R_{w}\rangle$ and $\langle R_{s} \rangle$
  along the centreline for the single square grid.}
\label{f:fig8}
\end{figure}

%MORE STUFF
In Figure \ref{f:fig8} we plot the streamwise evolution of $\langle
Q_{w}\rangle$, $\langle Q_{s}\rangle$ as functions of $x/x_{*}$ as
well as $\langle R_{w}\rangle$, $\langle R_{s} \rangle$ along the
centreline for the three simulations. Note that $\langle . \rangle$
means an average in time for a particular point in space. The first
important result is that $\langle Q \rangle=\langle
Q_{w}\rangle+\langle Q_{s}\rangle$ and $\langle R \rangle=\langle
R_{w}\rangle +\langle R_{s} \rangle$ along the centreline of the flow
are very close to zero for the three simulations, as expected in
homogeneous isotropic turbulence. This is not a trivial result as
already stated by \cite{laizetetal13} because grid-generated
turbulence is not homogeneous just downstream of the grid in the
production region where the four wakes are mixing together.
%It means that for our single square grid the production rate
%of strain, namely $R_{s}$ is on average equal to the production rate
%of enstrophy, namely $-R_{w}$, even in the production region where the
%flow is not homogeneous. Note that for a forward cascade from the
%large scale to the small scales, both $R_{s}$ and $-R_{w}$ in average
%have to be positive \cite{tsinober09}, which is the case in the
%present numerical data. $Q_s$ is always negative and is related to the
%dissipation $\varepsilon$. $Q_w$ is always positive and is directly
%related to the enstrophy. Figure \ref{f:fig8} is showing that in
%average, there is a local balance between dissipation and vorticity
%strength.  As a consequence, regions with strong vorticity, can be
%defined as those with high positive values of $Q$, while regions with
%high values of dissipation can be associated with strong negative
%values of $Q$.
The plots in Figure \ref{f:fig8} for the $SSG$-$HR$ and $SSG$-$LR$
simulations are very similar, the only difference being the small
shift for the peak of the plotted quantities that is slightly earlier
in the case of the $SSG$-$LR$ simulation. For the simulation
$SSG$-$ULR$ both the location and the intensity of the peak are
impacted by the poor resolution. As already observed by
\cite{laizetetal13} for a fractal square grid, $<Q_{w}>$ and $<R_{w}>$
are very close to zero between $x/x_* =0$ and $x/x_* = 0.3$. This can
be observed for the three simulations. The location of the first
non-zero values of average enstrophy and enstrophy production rates is
however different for the three simulations. This can be related to
the pile-up of energy at the small scales due to the low resolution,
resulting in spurious numerical enstrophy where the flow should be
irrotational.

\begin{figure}
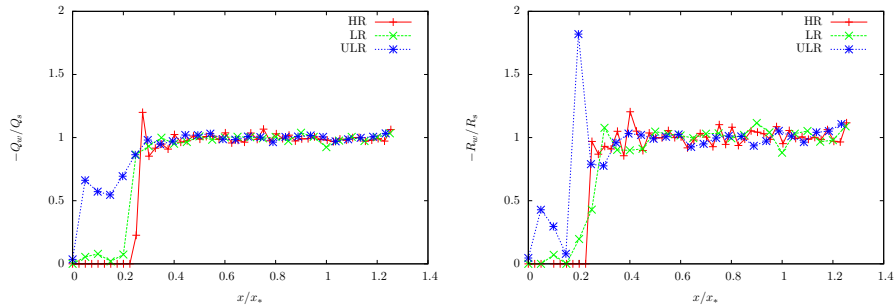

\centering
\resizebox{0.49\textwidth}{!}{\input{QQ_evol_ssg}}
\resizebox{0.49\textwidth}{!}{\input{RR_evol_ssg}}
\caption{Streamwise evolution of $\langle -Q_w/Q_s\rangle$ and
  $\langle -R_w/R_s\rangle$ along the centreline for the three
  simulations}
\label{f:fig9}
\end{figure}

 In order to better investigate the behaviour of the flow in the
 region $0< x/x_{*} < 0.3$, we plot in figure \ref{f:fig9} the
 streamwise evolution of $\langle -Q_w/Q_s\rangle$ and of $\langle
 -R_w/R_s\rangle$ on the centreline of the flow. The effect of the
 resolution can clearly be seen very close to the grid where the
 ratios $\langle -Q_w/Q_s\rangle$ and $\langle -R_w/R_s\rangle$ are
 virtually zero for the simulation with the highest resolution whereas
 they are clearly non-zero for the two other simulations. Based on the
 $SSG$-$HR$ simulation, we can say that $Q_s$ and $R_s$ are much
 larger than $Q_w$ and $R_w$ respectively, meaning that $Q_w$ should
 be virtually zero in the region $0 < x/x_* < 0.3$ along the
 centreline. It is a confirmation that the pile-up of energy for the
 simulations at low resolutions is creating numerical spurious
 enstrophy which can be seen in $Q_w$ and $R_w$. After $0.3x_*$, the
 three simulations are giving the same result with $\langle
 -Q_w/Q_s\rangle \approx \langle -R_w/R_s\rangle \approx 1$.

\begin{figure}
\centering
%\resizebox{0.45\textwidth}{!}{\input{QR_ULR_0-08}}
%\resizebox{0.45\textwidth}{!}{\input{QR_LR_0-08}}
%\resizebox{0.45\textwidth}{!}{\input{QR_HR_0-08}}
\includegraphics[angle=-0,width=0.99\textwidth]{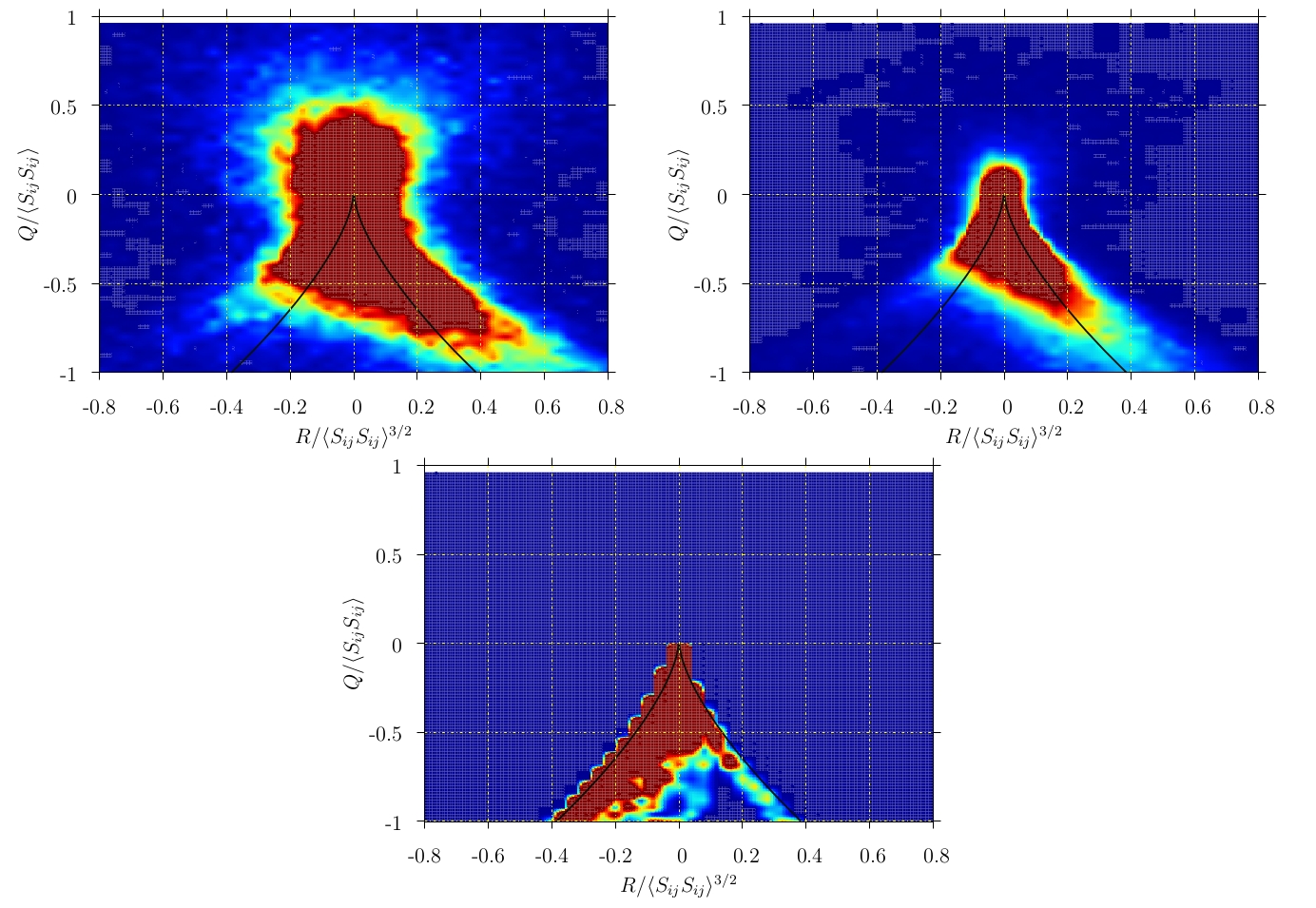}
\caption{Joint probability density function for the $Q$-$R$ diagram
  obtained at $x=0.08x_*$ for the $SSG$-$ULR$ (top left), $SSG$-$LR$
  (top right) and $SSG$-$HR$ (bottom) simulations.}
\label{f:fig10}
\end{figure}

\begin{figure}
\centering
%\resizebox{0.45\textwidth}{!}{\input{QR_ULR_0-2}}
%\resizebox{0.45\textwidth}{!}{\input{QR_LR_0-2}}
%\resizebox{0.45\textwidth}{!}{\input{QR_HR_0-2}}
\includegraphics[angle=-0,width=0.99\textwidth]{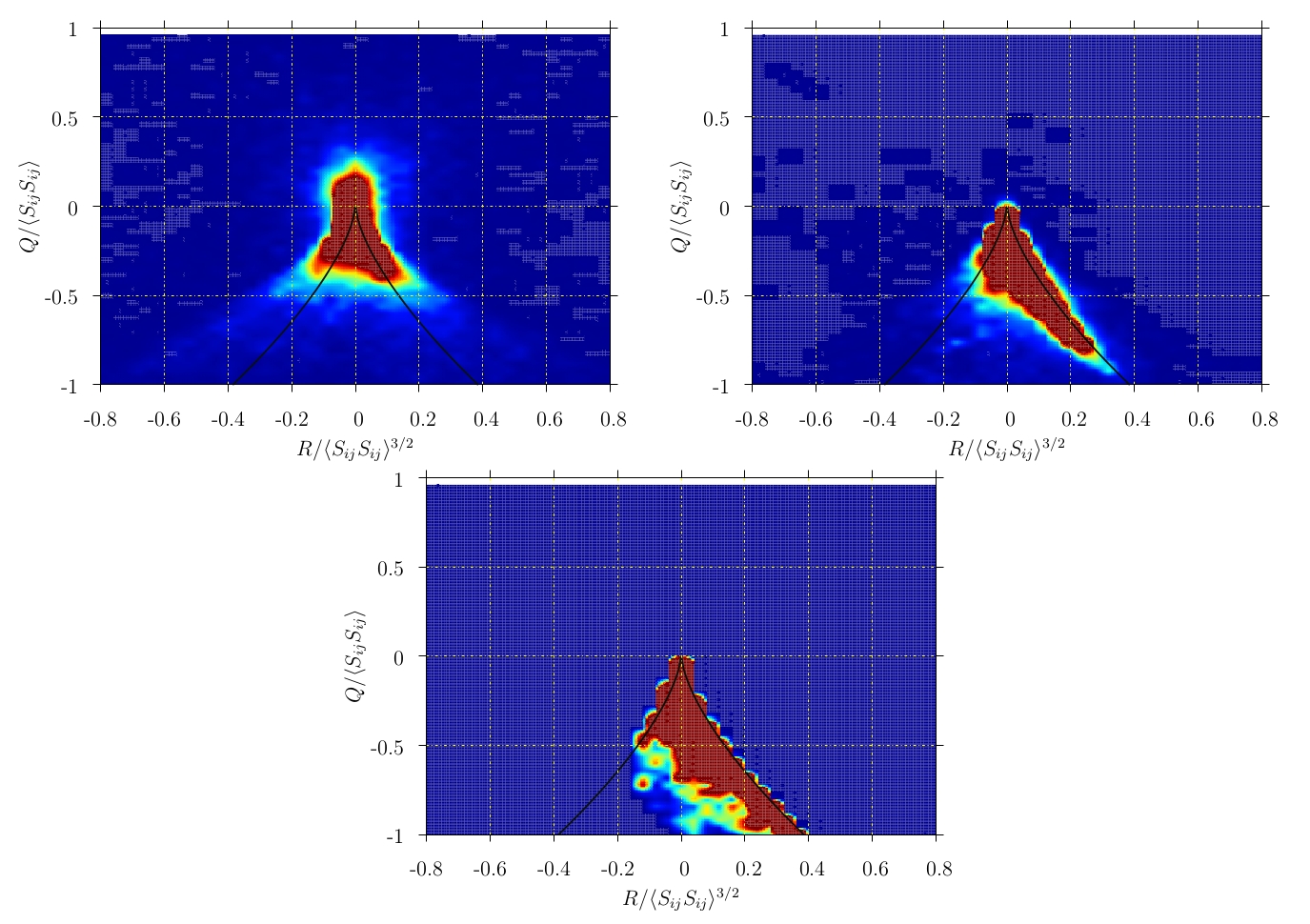}
\caption{Joint probability density function for the $Q$-$R$ diagram
  obtained at $x=0.2x_*$ for the $SSG$-$ULR$ (top left), $SSG$-$LR$
  (top right) and $SSG$-$HR$ (bottom) simulations.}
\label{f:fig11}
\end{figure}

\begin{figure}
\centering
%\resizebox{0.45\textwidth}{!}{\input{QR_ULR_0-5}}
%\resizebox{0.45\textwidth}{!}{\input{QR_LR_0-5}}
%\resizebox{0.45\textwidth}{!}{\input{QR_HR_0-5}}
\includegraphics[angle=-0,width=0.99\textwidth]{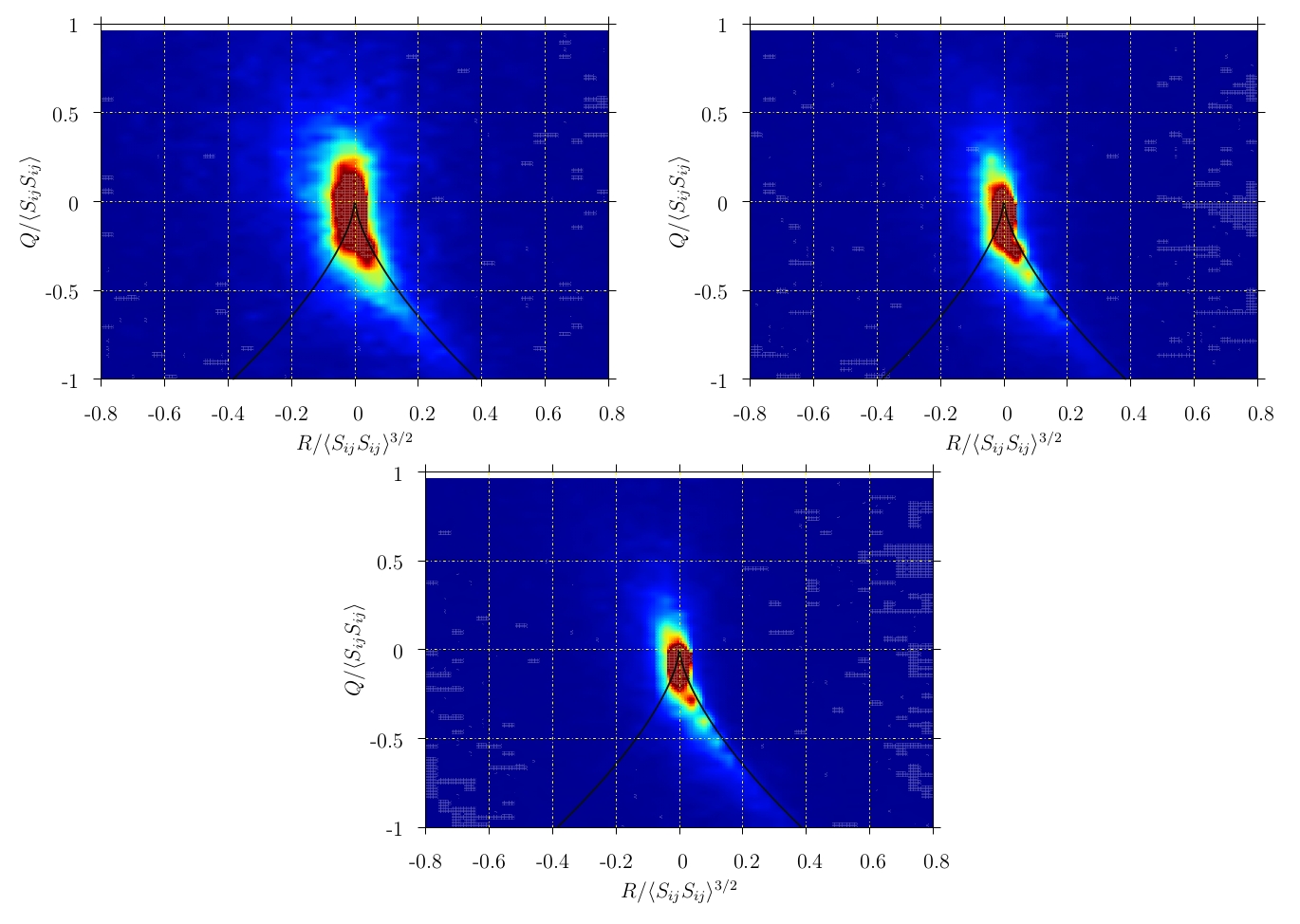}
\caption{Joint probability density function for the $Q$-$R$ diagram
  obtained at $x=0.5x_*$ for the $SSG$-$ULR$ (top left), $SSG$-$LR$
  (top right) and $SSG$-$HR$ (bottom) simulations.}
\label{f:fig12}
\end{figure}

\begin{figure}
\centering 
%\resizebox{0.45\textwidth}{!}{\input{QR_ULR_1-0}}
%\resizebox{0.45\textwidth}{!}{\input{QR_LR_1-0}}
%\resizebox{0.45\textwidth}{!}{\input{QR_HR_1-0}}
\includegraphics[angle=-0,width=0.99\textwidth]{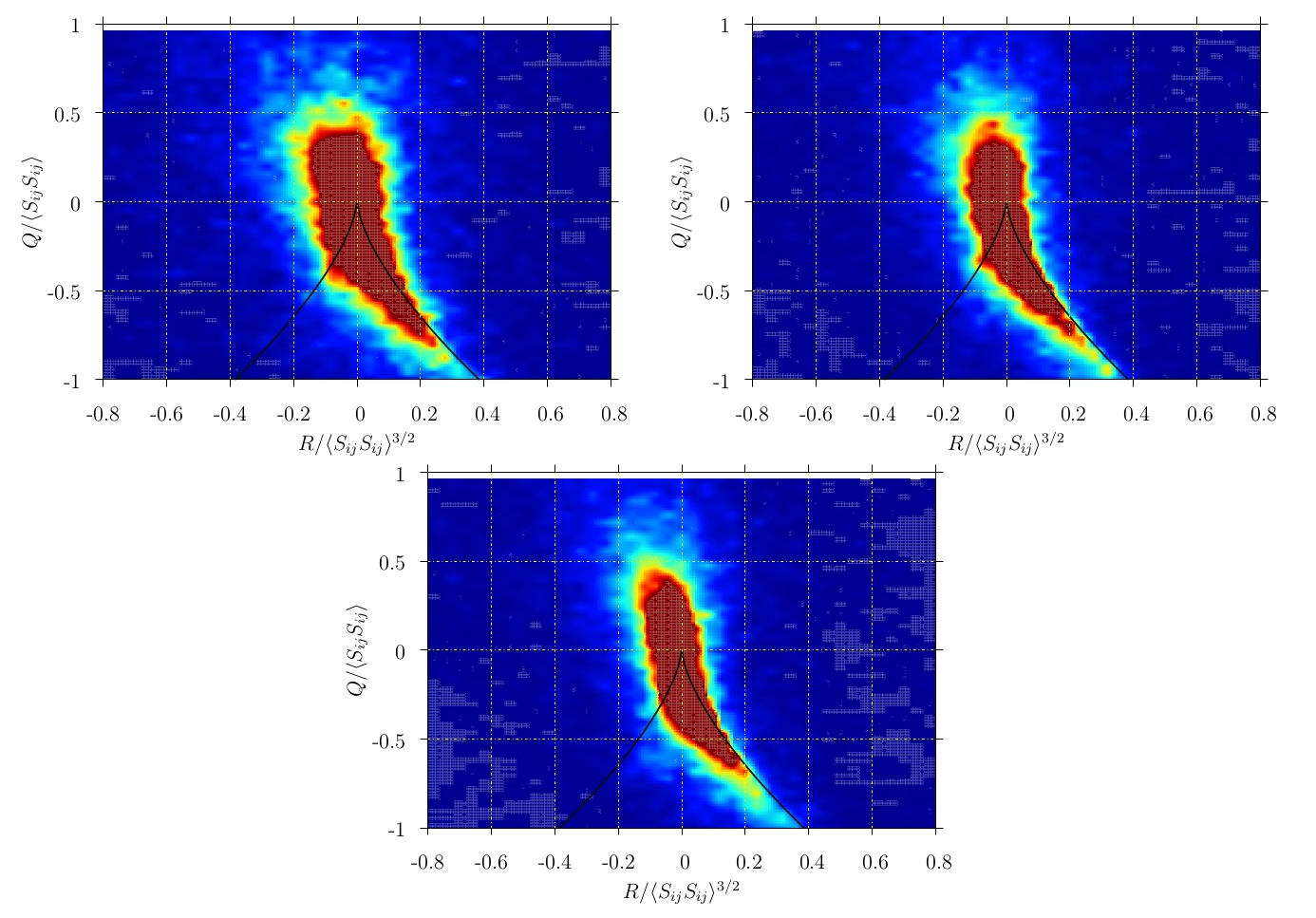}
\caption{Joint probability density function for the $Q$-$R$ diagram
  obtained at $x=x_*$ for the $SSG$-$ULR$ (top left), $SSG$-$LR$ (top
  right) and $SSG$-$HR$ (bottom) simulations.}
\label{f:fig13}
\end{figure}

It is of interest to see how the $Q$-$R$ diagram is evolving
downstream of the single square grid and how it is affected by the
resolution of the simulations. The plots presented in figure
\ref{f:fig10}, \ref{f:fig11}, \ref{f:fig12} and \ref{f:fig13} are
obtained for four streamwise locations corresponding to
$x=0.08x_*,~0.2x_*,0.5x_*$ and $x_*$ and are based on data collected
in time over a period equivalent to $T=16sec$. As suspected, there is
a clear difference very close to the grid between the three
simulations as shown in figure \ref{f:fig10} at $x=0.08x_*$. Based on
the simulation with the best resolution, the flow should be dominated
by flow regions where $R<0$ and $Q<0$, as already observed by
\cite{zhouetal14} in a very similar flow configuration. The trend is
less obvious with the intermediate resolution whereas the simulation
with the lowest resolution is producing a completely different $Q$-$R$
diagram with a tear drop shape for $Q<0$ and $R>0$.

Further downstream, at $x=0.2x_*$, the $Q$-$R$ diagram is quite
different with a flow dominated by flow regions where $R>0$ and $Q<0$
as shown in figure \ref{f:fig11}.  It should be noted that $Q$ is
almost always negative at this point which shows that there is still
no enstrophy at this streamwise location. The $Q$-$R$ diagram obtained
here is very similar to the ones obtained by \cite{dasilva&pereira08}
in the non rotational region surrounding a spatially evolving
turbulent jet. The $SSG$-$HR$ and $SSG$-$LR$ simulations are in fairly
good agreement with each other for this location. The simulation with
the lowest resolution is producing a nearly symmetric $Q$-$R$ diagram,
meaning that it is not possible to track any fluid flow dynamics at
this resolution using the $Q$-$R$ diagram. It is consistent with the
pile-up of energy at the small scales, altering the flow motion at
this location. We will see later that this nearly symmetric shape for
the $Q$-$R$ diagram is also the signature of a random white noise field
\cite{tsinober09}.

For $x=0.5x_*$, we can see in figure \ref{f:fig12} that the $Q$-$R$
diagram is at the beginning of adopting its usual tear drop shape. At
this location, the effect of the resolution is less pronounced, the
only difference being the size of the dark red region which is slightly
larger when the resolution is decreased. Note that this streamwise
location is in the decay region for the turbulence, just after the
peak shown in figure \ref{f:fig3}.

Finally, further downstream in the decay region for $x=x_*$, it can be
seen in figure \ref{f:fig13} that the $Q$-$R$ diagram has a tear drop
shape and that the resolution is not damaging this tear drop
shape. This suggests that in this region, the pile-up of energy at the
small scales is not affecting the flow motion, at least not enough to
strongly impact the $Q$-$R$ diagram.

\begin{figure}
\centering
\resizebox{0.6\textwidth}{!}{\input{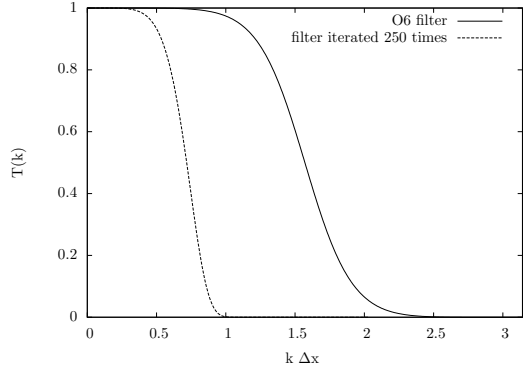}}
\caption{Filtering transfer function $T(k)$ versus wave number $k$ for
  the sixth-order compact operator (\ref{filter}) used in this work.}
\label{f:fil}
\end{figure}

In order to better understand the effect of the small-scale pile-up of
energy on the $Q$-$R$ diagram, we are now going to filter the data
where the pile-up of energy is damaging the $Q$-$R$ diagram and see if
it is possible to recover the diagram obtained with the simulation
with the highest resolution.  The filtering procedure is based on the
sixth-order compact operator proposed by \cite{lele92}
\begin{equation}
\frac{3}{10} \hat{f}_{i-2} + \hat{f}_i + \frac{3}{10} \hat{f}_{i+2}
= \frac{1}{2} f_i+ \frac{3}{8} (f_{i+1}-f_{i-1}) +
\frac{3}{20} (f_{i+2}-f_{i-2}) +  \frac{1}{40} (f_{i+3}-f_{i-3})
\label{filter}
\end{equation}
with $f_i=f(x_i)$, $\hat{f}_i=\hat{f}(x_i)$ and $x_i=(i-1)\Delta x$
for $(i=1,...,n_x)$, all those quantities being defined for
$[0,L_x]$. $\hat{f}(x_i)$ corresponds to the filtered quantity.  The
associated filtering transfer function $T(k)$ is 
\begin{equation}
T(k)=\frac{1/2+(3/4)\cos(k)+(3/10)\cos(k)+(1/20)\cos(k)}{1+(3/5)\cos(2k)}
\end{equation}
and is plotted in Figure \ref{f:fil}. The filter operator, applied in
the three spatial directions on the three components of the velocity,
can be see as a low pass filter for which the filtering effect is
confined to the shortest wavelengths. In the present work, the filter
operator is applied 250 times to each 3D snapshot of the $SSG$-$LR$
simulation, corresponding to the elimination of the smallest scales of
the flow up to $10 \eta$. It is necessary to apply the filter operator
250 times in order to have clean 3D snapshots free of any numerical
oscillations. As shown in Figure \ref{f:fil}, it corresponds to a
filter width of about $3.5\Delta x$.  Then the $Q$-$R$ diagram is
produced from the filtered data and is compared with the $Q$-$R$
diagram obtained from the $SSG$-$HR$ simulation at the same
location. Note that the $Q$-$R$ diagrams are not computed in time any
more. Each $Q$-$R$ diagram presented in figures 15 to 19 is obtained
from a single snapshot in a small 3D cube $0.025x_*\times 0.025x_*
\times 0.025x_*$ around a specific streamwise location on the
centreline. We have checked over our different uncorrelated snaphots
that the data presented in this study are representative of the flow
dynamics for a given streamwise location.

\begin{figure}
\centering
%\resizebox{0.45\textwidth}{!}{\input{QR_S_HR_00-050}}
%\resizebox{0.45\textwidth}{!}{\input{QR_S_HR_10-050}}
%\resizebox{0.45\textwidth}{!}{\input{QR_S_HR_11-050}}
\includegraphics[angle=-0,width=0.99\textwidth]{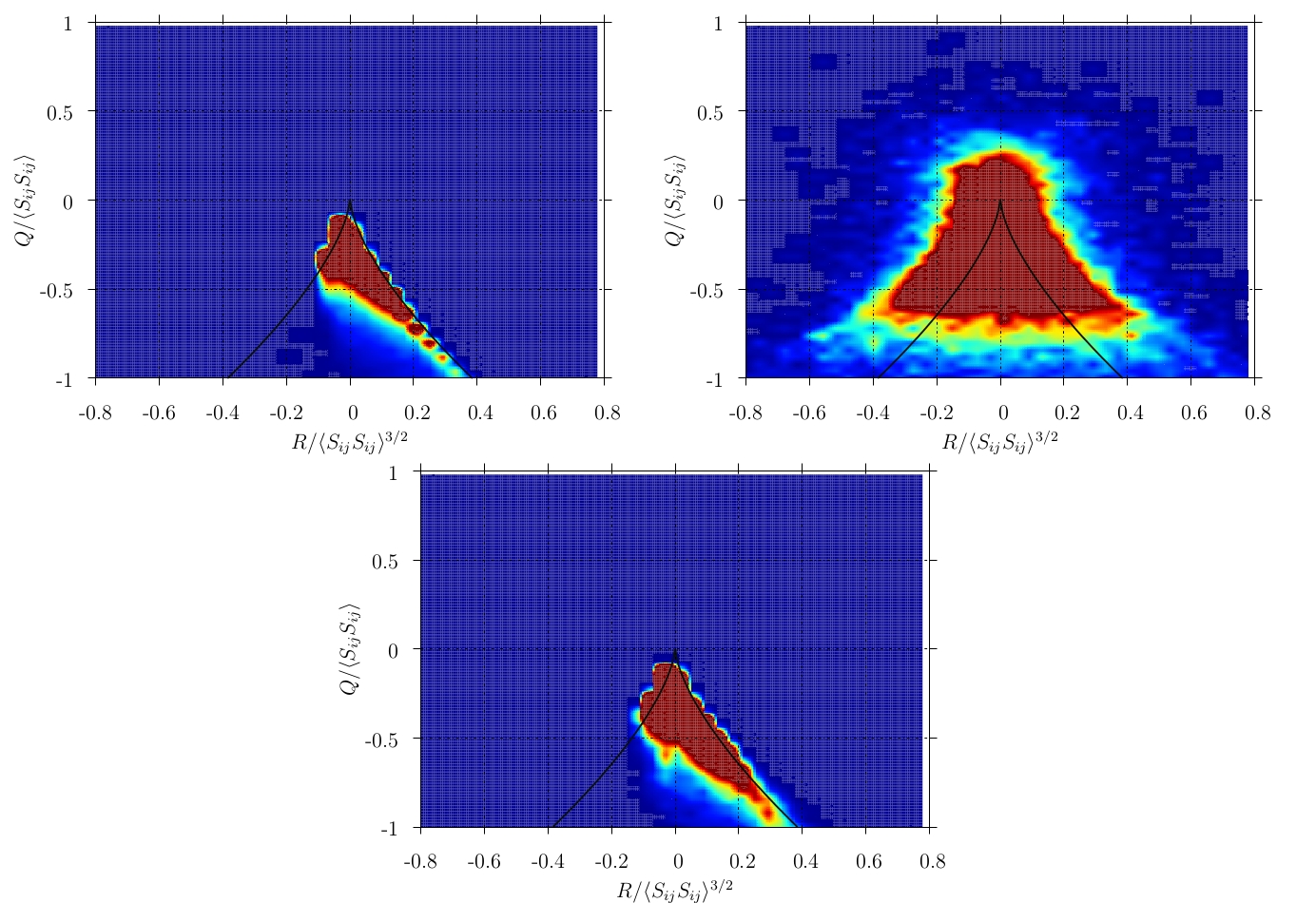}
\caption{Joint probability density function for the $Q$-$R$ diagrams
  obtained from the $SSG$-$HR$ simulation at $x=0.2x_*$ for a
  ($0.025x_*\times 0.025x_* \times 0.025x_*$) cube (125,000 mesh
  nodes) with no added random noise and no filtering (top left), with
  added random noise (top right) and with added random noise and
  filtering (bottom).}
\label{f:fig14}
\end{figure}

First, to validate our filtering procedure, we take a snapshot from
the $SSG$-$HR$ simulation, superimpose a random white noise to it to
mimic the numerical oscillations due to the pile-up of energy at the
small scales and then filter the altered snapshot with the aim to
recover the $Q$-$R$ diagram produced by the clean snapshot.  In figure
\ref{f:fig14}, three $Q$-$R$ diagrams are presented. They are obtained
for the $SSG$-$HR$ simulation at $x=0.2x_*$. The first important
result is that the $Q$-$R$ diagram shown in figure \ref{f:fig14} (top
left) is very similar to the one obtained at the same location with
the data in time (see figure \ref{f:fig11} bottom). The $Q$-$R$
diagram presented in figure \ref{f:fig14} (top right) is obtained with
a random white noise superimposed to the velocity field. This noise
corresponds to a $0.025\%$ uncertainty in the mean value of
$U_{\infty}$. The shape of the $Q$-$R$ diagram is perfectly symmetric
and is very similar to the one obtained for the $SSG-ULR$ simulation
at $x=0.2x_*$ (see figure \ref{f:fig11} top left).  It means that the
pile-up of energy at the small scales and the random noise are
damaging  the $Q$-$R$ diagram in a similar fashion. Furthermore, for
this particular location, the $Q$-$R$ diagram is clearly dominated by
the added random noise (or by the low resolution). When the filtering
procedure is applied to the data with a superimposed random white
noise, the $Q$-$R$ diagram observed in figure \ref{f:fig14} (bottom)
is very similar to the one obtained for the raw data.

\begin{figure}
\centering
%\resizebox{0.45\textwidth}{!}{\input{QR_S_HR_00-600}}
%\resizebox{0.45\textwidth}{!}{\input{QR_S_HR_10-600}}
%\resizebox{0.45\textwidth}{!}{\input{QR_S_HR_11-600}}
\includegraphics[angle=-0,width=0.99\textwidth]{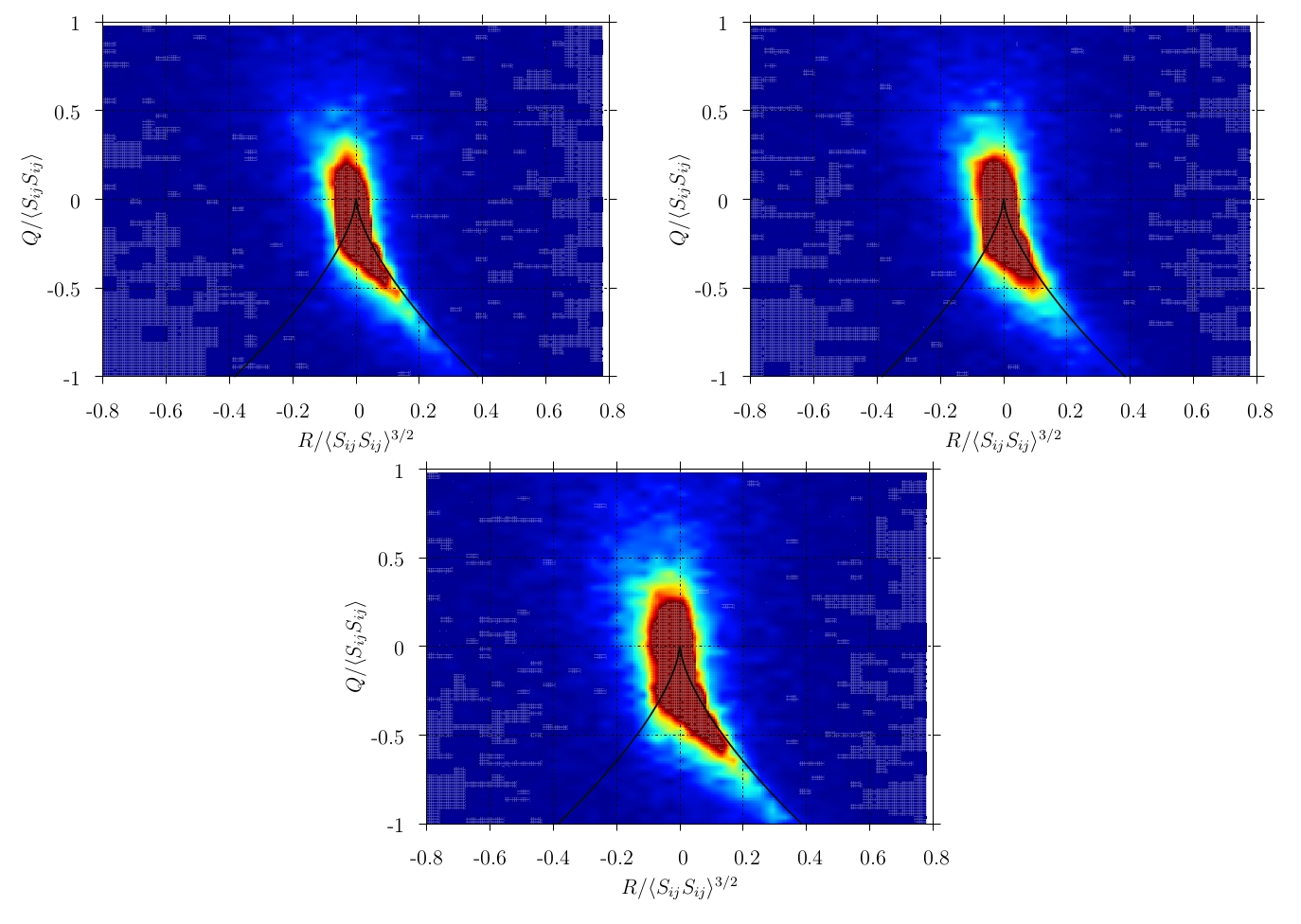}
\caption{Joint probability density function for the $Q$-$R$ diagrams
  obtained from the $SSG$-$HR$ simulation at $x=0.6x_*$ for a
  ($0.025x_*\times 0.025x_* \times 0.025x_*$) cube (125,000 mesh
  nodes) with no added random noise and no filtering (top left), with
  added random noise (top right) and with added random noise and
  filtering (bottom).}
\label{f:fig15}
\end{figure}

The same procedure is repeated downstream of the grid at $x=0.6x_*$. The data
are presented in figure \ref{f:fig15}. It can be seen that the
addition of a random white noise has only a limited impact on the
$Q$-$R$ diagram, the only visible difference being a reduced tear drop
tail and a broadening of the dark red region, as already observed by
\cite{buxtonetal11ef} in a turbulent mixing layer flow. As expected,
the effect of the filtering procedure on the data with a superimposed
random white noise is also quite limited, suggesting that for this
particular location, the $Q$-$R$ diagram is mainly dominated by large
scale structure features. The alteration of the small scales at this
streamwise location seems to be quite limited on strain-rate and
rotation tensors. It is worth pointing out that there is a big
difference between adding spurious noise to a well-resolved set of
data and having spurious numerical artefacts in an under-resolved DNS
in which the dynamics among all the scales maybe incorrectly
reproduced. For this numerical investigation, we are just trying to
find a way to filter under-resolved DNS data to better understand how
the resolution is affecting the strain-rate and rotation tensors.

\begin{figure}
\centering
%\resizebox{0.45\textwidth}{!}{\input{XL200-100ULR}}
%\resizebox{0.45\textwidth}{!}{\input{XL211-100ULR}}
\includegraphics[angle=-0,width=0.99\textwidth]{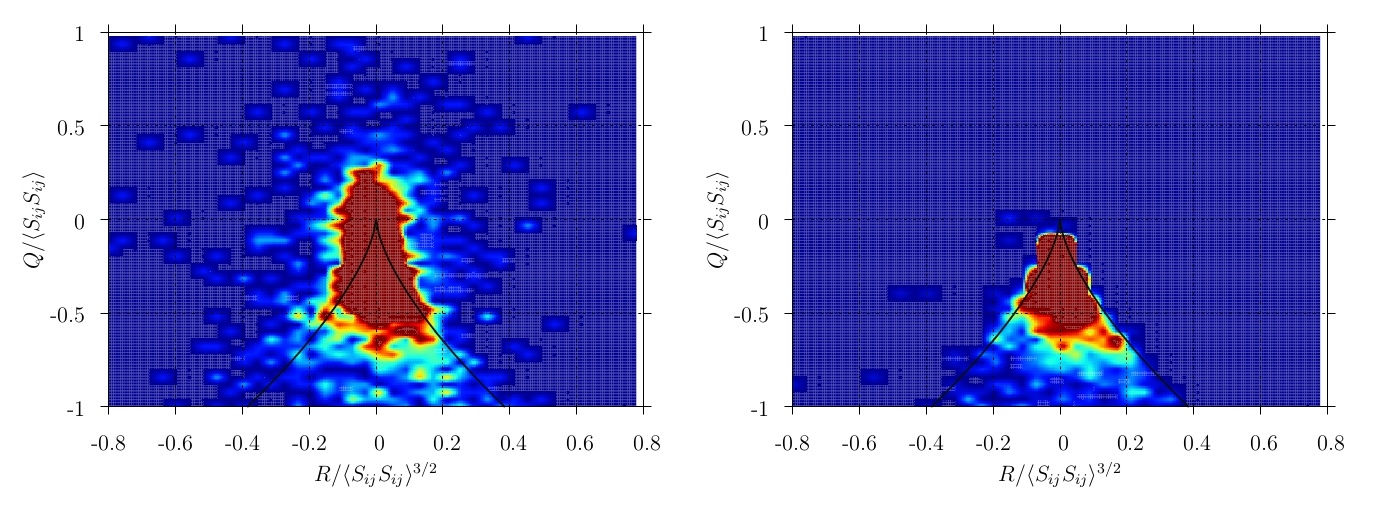}
\caption{Joint probability density function for the $Q$-$R$ diagrams
  obtained from the $SSG$-$LR$ simulation at $x=0.2x_*$ for a
  ($0.025x_*\times 0.025x_* \times 0.025x_*$) cube (15,625 mesh nodes)
  with no filtering (left) and with filtering (right).}
\label{f:fig16}
\end{figure}

The filtering procedure is now applied to a set of snapshots obtained
from the $SSG$-$LR$ simulation where spurious numerical artefacts are
present. Figure \ref{f:fig16} shows the $Q$-$R$ diagram obtained
before and after the filtering procedure for a ($0.025x_*\times
0.025x_* \times 0.025x_*$) cube (15,625 mesh nodes) located at
$x=0.2x_*$ on the centreline. We can see that the non-filtered
snapshots are producing a symmetric shape for the $Q$-$R$ diagram with
positive values for $Q$, signature of spurious enstrophy caused by the
pile-up of energy at the small scales. When the data are filtered, all
the positive values of $Q$ are removed and the $Q$-$R$ diagram has a
similar shape to the one obtained with the $SSG$-$HR$ simulation for
the same location, with negative values of $Q$ weakly skewed toward
$R>0$.

\begin{figure}
\centering
%\resizebox{0.45\textwidth}{!}{\input{XL200-320ULR}}
%\resizebox{0.45\textwidth}{!}{\input{XL211-320ULR}}
\includegraphics[angle=-0,width=0.99\textwidth]{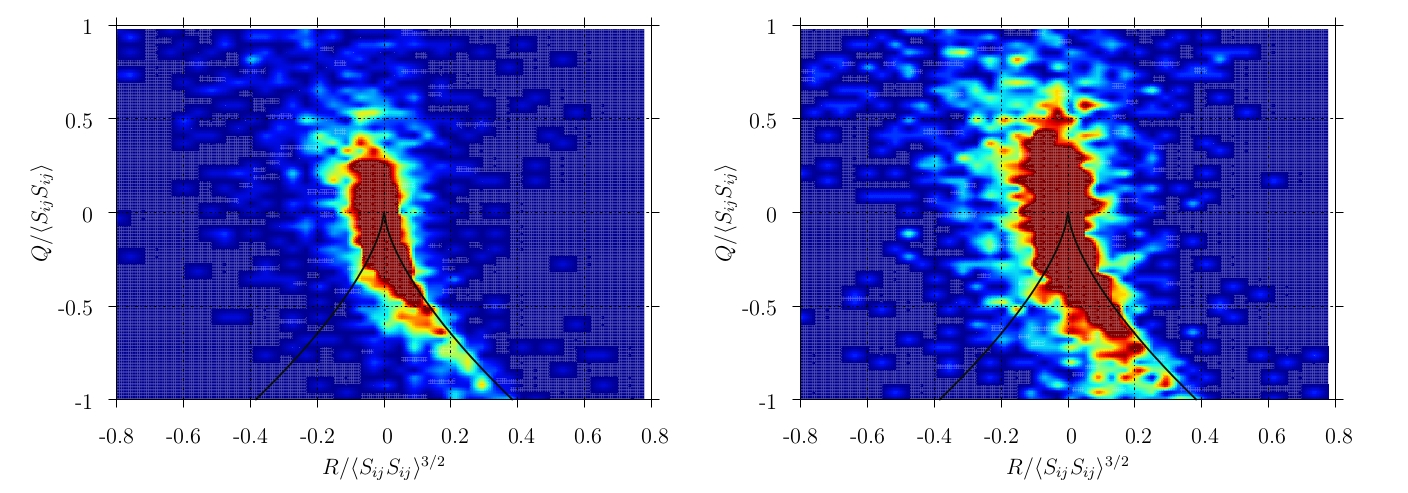}
\caption{Joint probability density function for the $Q$-$R$ diagrams
  obtained from the $SSG$-$LR$ simulation at $x=0.6x_*$ for a
  ($0.025x_*\times 0.025x_* \times 0.025x_*$) cube (15,625 mesh nodes)
  with no filtering (left) and with filtering (right).}
\label{f:fig17}
\end{figure}

Further downstream for $x=0.6x_*$, the usual tear drop shape can be
observed for both the non-filtered and filtered data, as shown in
figure \ref{f:fig17}. Like previously observed for the 
$SSG$-$HR$ simulation for which random white noise was added, it seems that
the spurious numerical errors for the small scales are not impacting
too much the shape of the $Q$-$R$ diagram. The main difference between
the filtered data and the non-filtered data is the size of the dark
region which is larger when the data are filtered, with more data
points for $Q>0$. It is a notable result, suggesting that an
under-resolved DNS can qualitatively predict the behaviour of the
strain-rate and rotation tensors at least when the flow is dominated
by large scale features.

\section{Conclusion}
Direct Numerical Simulations of the turbulence generated by a single
square grid have been presented in this paper in order to investigate
the influence of the spatial resolution on fine-scale features and in
particular on the strain-rate and rotation tensors. Careful
comparisons with hot-wire experiments have been carried out on the
centreline of the flow for first, second, third and fourth order
moments of one-point flow velocities. For those quantities, we show
that even the simulation with the lowest resolution ($\Delta x$ at
worse equal to $7\eta$, at best equal to $2\eta$) is able to
reproduce the experimental results within an error margin of about
$10\%$. For the third and fourth order moments, it seems that the
numerical data are not converged enough in time and the quality of the
present numerical data would be greatly improved by increasing the
level of convergence by one order of magnitude or two.  For the first
and second order moments, a resolution of $\Delta x \approx 5 \eta$
seems to be enough to match experimental data within a margin of
$5\%$.

Concerning the $Q-R$ diagram and the strain-rate and rotation tensors,
the results are strongly dependent on both the resolution and the
streamwise location. In the production region, upstream of the peak of
turbulence, the flow is dominated by strain ($Q<0$ for the simulation
$DNS-HR$) and the resolution is deeply impacting the small-scale
features of the flow with positive values for $Q$ through the addition
of spurious numerical artefacts when the spatial resolution is worse
than $4\eta$, at least for our code \verb|Incompact3d|, based on
sixth-order finite-difference schemes on a Cartesian mesh. The $Q-R$
diagram can be used in our code as an indicator of the presence in the
flow of non-physical features. The influence of our numerical
artefacts on the $Q-R$ are similar to a random white noise. In the
decay region, where the usual tear drop shape is observed for all the
simulations, it is more difficult to quantify the influence of
spurious numerical artefacts using the $Q-R$ diagram. The only
noticeable difference is an increase of the size of the $Q-R$ diagram
when the spatial resolution is decreased. The conclusion is that it is
necessary to have a very fine spatial resolution of less than $2\eta$
for a correct reproduction of the strain-rate and rotation tensors.

%Finally, a notable result observed in this numerical study is that the
%present results are suggesting that the new dissipation law
%\cite{vassilicos15}, for which $C_{\epsilon}$ is not constant, is only
%valid above a certain value for $Re_{\lambda}$, value which is
%grid-dependant. We found that for low inflow velocities (the DNS and
%the experiment with $2.5m/s$), $Re_{\lambda}$ in the decay region of
%the flow is not decreasing as expected but is more or less constant
%along the centreline. It seems that for this regime, the flow is very
%similar to a turbulent jet for which the rms of the fluctuating
%velocity $u^{\prime}$ is decaying as $1/x$ whereas the Taylor
%micro-scale is increasing as $x$ resulting in a constant behaviour for
%$Re_{\lambda}$.

\section*{Acknowledgements}

We are grateful to Eric Lamballais for kindly commenting on an early
draft of the manuscript.  We also acknowledge EPSRC Research grant
EP/L000261/1 for access to UK supercom- puting resources and PRACE for
awarding us access to resource SUPERMUC based in Germany at
Leibniz-Rechenzentrum (Leibniz Supercomputing Centre). The authors
were supported by an ERC Advanced Grant (2013-2018) awarded to
J.C.Vassilicos.

\end{document}